\documentclass[preprint2,tighten,times,appendixfloats]{aastex6}

\usepackage{graphicx}
\usepackage{xspace}
\usepackage{natbib}
\usepackage{microtype}
\usepackage{txfonts}

\newcommand{\ns}{\ensuremath{\mathrm{s}_{1/2}}}
\newcommand{\nsf}{\ensuremath{\mathrm{s}^2}}
\newcommand{\np}[1][\empty]{\ensuremath{\mathrm{p}^{#1}_{1/2}}}
\newcommand{\npp}[1][\empty]{\ensuremath{\mathrm{p}^{#1}_{3/2}}}
\newcommand{\npf}{\ensuremath{\mathrm{p}^6}}
\newcommand{\sle}[2]{\ensuremath{^{#1}\mathrm{#2}}}
\newcommand{\slo}[2]{\ensuremath{^{#1}\mathrm{#2}^{\mathrm{o}}}}
\newcommand{\slj}[3]{\ensuremath{^{#1}\mathrm{#2}_{#3}}}
\newcommand{\sljo}[3]{\ensuremath{^{#1}\mathrm{#2}^{\mathrm{o}}_{#3}}}
\newcommand{\lss}[1][\empty]{\ensuremath{\mathrm{s}^{#1}}}
\newcommand{\lsp}[1][\empty]{\ensuremath{\mathrm{p}^{#1}}}

\newcommand{\smalltablefont}{\fontsize{5.7}{7}\selectfont}

\begin{document}

\title{\vspace*{-2mm}Laboratory Measurements of the K-shell transition energies in
L-shell ions of Si and S}

\shorttitle{K$\alpha$ Lines of Si and S}
\shortauthors{Hell et al.}

\author{ N.~Hell\altaffilmark{1,2},
  G.V.~Brown\altaffilmark{2},
  J.~Wilms\altaffilmark{1},
  V.~Grinberg\altaffilmark{3},
  J.~Clementson\altaffilmark{2},
  D.~Liedahl\altaffilmark{2},
  F.S.~Porter\altaffilmark{4},
  R.L.~Kelley\altaffilmark{4},
  C.A.~Kilbourne\altaffilmark{4},
  P.~Beiersdorfer\altaffilmark{2}
}

\altaffiltext{1}{Dr.~Karl Remeis-Sternwarte and Erlangen Centre for
  Astroparticle Physics, Universit\"at Erlangen-N\"urnberg,
  Sternwartstr.~7, 96049 Bamberg, Germany
  \email{natalie.hell@sternwarte.uni-erlangen.de}} 

\altaffiltext{2}{Lawrence Livermore National Laboratory, 7000 East
  Ave., Livermore, CA 94550, USA} 

\altaffiltext{3}{Massachusetts Institute of Technology, Kavli
  Institute for Astrophysics and Space Research, Cambridge, MA 02139,
  USA} 

\altaffiltext{4}{NASA-GSFC, 8800 Greenbelt Road, Greenbelt, MD 20771,
  USA} 

\submitted{ApJ, in press}

\begin{abstract}
We have measured the energies of the strongest 1s--2$\ell\
(\ell=\text{s,p})$ transitions in He- through Ne-like silicon and
sulfur ions to an accuracy of $<1\,\mathrm{eV}$ using Lawrence
Livermore National Laboratory's electron beam ion traps, EBIT-I and
SuperEBIT, and the NASA/GSFC EBIT Calorimeter Spectrometer (ECS). We
identify and measure the energies of 18 and 21 X-ray features from
silicon and sulfur, respectively. The results are compared to new
Flexible Atomic Code calculations and to semi-relativistic Hartree
Fock calculations by Palmeri et al.\ (2008). These results will be
especially useful for wind diagnostics in high mass X-ray binaries,
such as Vela X-1 and Cygnus X-1, where high-resolution spectral
measurements using \textsl{Chandra}'s high energy transmission grating
has made it possible to measure Doppler shifts of
$100\,\mathrm{km\,s}^{-1}$. The accuracy of our measurements is
consistent with that needed to analyze \textsl{Chandra} observations,
exceeding \textsl{Chandra}'s $100\,\mathrm{km\,s}^{-1}$ limit. Hence,
the results presented here not only provide benchmarks for theory, but
also accurate rest energies that can be used to determine the bulk
motion of material in astrophysical sources. We show the usefulness of
our results by applying them to redetermine Doppler shifts from
\textsl{Chandra} observations of Vela X-1. 
\end{abstract}

\section{Introduction}

Prominent absorption and emission X-ray features from highly charged
silicon and sulfur ions have been detected and measured in a medley of
celestial sources, including solar flares \citep{neupert1971a}, other
stellar coronae \citep[e.g.,][]{kastner2002a,huenemoerder2013a},
various types of Active Galactic Nuclei \citep[e.g.,][]{leejc2001a,
  kaspi2002a,kinkhabwala:2002fx,holczer2007a,holczer2012a,reeves2013a},
and high-mass X-ray binaries
\citep[HMXB; e.g.,][]{sako02a,boroson2003a,watanabe06a, chang07a,
hanke08a,miskovicova:16a}. HMXBs, although well studied and
cataloged, are not yet fully understood. In general, they consist of
a massive O- or B-type star in orbit with a compact object, either a
black hole or neutron star. X-ray emission or absorption features from
these sources are generated when the luminous
($10^{36}\ldots10^{38}\,\mathrm{erg}\,\mathrm{s}^{-1}$) X-ray
continuum from the accreting compact object irradiates, ionizes, and
fluoresces the stellar wind material ejected from the companion star.
Because the stellar wind of the massive companion is radiation driven,
the ionizing nature of the X-ray continuum affects not only the wind
structure, but also the mass loss rate of the companion star. Hence,
K$\alpha$ transitions originating in the wind have not only been used
to determine the ion structure and motion of the wind, but also
provide insight into the mass loss rate of the companion star and the
strength of the X-ray continuum. For example, in the case of
\object{Vela X-1}, \citet{sako02a}, \citet{schulz02a},
\citet{goldstein04a}, and \citet{watanabe06a} report high resolution
X-ray emission spectra from 2p$\rightarrow$1s, i.e., K$\alpha$,
transitions from both L- and K-shell silicon and sulfur ions.
\citet{sako02a} identify resolved line emission from O-like
\ion{Si}{7} through H-like \ion{Si}{14}, and an unresolved feature
identified as \ion{Si}{2}--\ion{Si}{6}. \citet{goldstein04a} find 
the motion of different ions of the same element to be non-uniform,
based on the limited quality of their used reference wavelengths.
\citet{watanabe06a} build a three dimensional Monte-Carlo radiative
transfer model and report a mass loss rate for the companion star and
the structure of the wind, although they do not analyze the line
emission from the L-shell silicon ions, but only from H-like Si. In
the case of \object{Cygnus X-1}, the K$\alpha$ absorption features in
L-and K-shell ions of silicon and sulfur have been measured and used
to diagnose the nature of the stellar wind
\citep{hanke08a,hell2013a,miskovicova:16a}. Specifically, these
features have been shown to be produced by ``clumps'' of
onion-structured material, where the inner layers are colder, denser,
and less ionized, moving in and out of the observational line
of sight.

In the case of multielectron L-shell ions of silicon and sulfur, the
utility of the associated X-ray line diagnostics is limited and often
precluded by the relatively poor accuracy of the atomic reference
data. Accurate calculations of the atomic structure of these ions is
challenging because correlation effects among multiple electrons must
be taken into account. Historically, Hartree-Fock calculations of
\citet{house69a} were used to interpret high resolution solar spectra
\citep{fritz1967a}, and more recently have been used to analyze data
from both Vela X-1 \citep{schulz02a,goldstein04a} and Cygnus X-1
\citep{hankediss}. However, \citet{house69a} only provide simplified
data listing only a single transition for each ion. To provide a more
complete and accurate data set, more sophisticated calculations have
been completed using more advanced atomic models. For example,
\citet{behar02a} used the Hebrew University Lawrence Livermore Atomic
Code \citep[HULLAC;][and references
therein]{klapisch1971a,klapisch2006a} to calculate transition energies
and line strengths for the strongest K-shell transitions in He-
through F-like silicon and sulfur ions. At present, the most complete
calculation is provided by \citet[][P08]{palmeri08a}, who use a
semi-relativistic Hartree-Fock code to calculate level energies,
transition wavelengths, and radiative decay rates for $\sim$1400
K-shell transitions in silicon and sulfur ions. The variation among
the inner-shell transition energies calculated with various codes is
$\sim$2--5\,eV, i.e., on the order of several
100\,$\mathrm{km}\,\mathrm{s}^{-1}$ for the diagnostically important
L-shell silicon K$\alpha$ lines. This variation is comparable to the
expected Doppler shift of the L-shell silicon K$\alpha$ lines
\citep{watanabe06a,liedahl08a,miller05a,miller12a,miskovicova:16a},
and significantly larger than the systematic wavelength error of
\textsl{Chandra}'s High Energy Transition Grating Spectrometer (HETGS),
which is on the order of $100\,\mathrm{km}\,\mathrm{s}^{-1}$
\citep{marshall2004a,canizares05a,chandrapog2015}. Hence, the main
systematic uncertainty in the determination of Doppler shifts from
X-ray lines is our knowledge of atomic physics. This has been pointed
out before in studies of the K-shell lines in L-shell oxygen ions
\citep{schmidt2004a,gu2005a}.
 
When comparing atomic databases commonly used to interpret both Solar
and extra-Solar X-ray spectra, the data from P08 are found in the
Universal Atomic DataBase (uaDB) accompanying XSTAR
\citep{bautista01a}; however, they are not included in either the
atomic physics for astrophysics database, AtomDB v2 \citep{foster12a}
or the CHIANTI atomic physics database \citep{dere97a,landi2013a}.
AtomDB v2 only includes K-shell transitions in helium-like and
hydrogen-like ions; CHIANTI only includes H-like, He-like, and Li-like
transitions.

There is one previous measurement available for L-shell transitions in
Be- through F-like Si and S ions. \citet{faenov94a} measured transitions
produced in a $\mathrm{CO}_2$ laser-produced plasma. They  also
provided a comparison to their own theoretical calculations. The
density of this plasma is significantly higher than typical densities
in an astrophysical environment. The spectra reported by
\citet{faenov94a} therefore comprise mainly dielectronic satellites
(see their Tables~I and~II) and are only of limited applicability for
our purpose. 

\enlargethispage{\baselineskip}

Here, we report results of measurements of the 1s--$2\ell\
(\ell=\mathrm{s, p})$ K-shell energies in He- to Ne-like ions of
silicon and sulfur in a coronal plasma produced with the Lawrence
Livermore National Laboratory electron beam ion traps
(Section~\ref{sec:measurement}). To gauge the systematic uncertainty
inherent to calculations of many-body atomic systems, we compare these
measurement results (Section~\ref{sec:fitmethod}) to line energy
calculations performed with two popular atomic codes used for line
identification, namely our own calculations with the Flexible Atomic
Code \citep[FAC;][]{gu04b,gu08a} (Section~\ref{sec:lineIDFAC}) and the
tables of P08 (Section~\ref{sec:facvspalm}). In addition, we list the
centers of major line blends as a reference for observations with
moderate resolution and derive new Doppler shifts for Vela~X-1 based
on our laboratory measured values (Section~\ref{sec:blends}). We
summarize our results in Section~\ref{sec:conclusion}.

\section{Measurement}\label{sec:measurement}

\subsection{Experimental Setup}
The measurements presented here were carried out using the Lawrence
Livermore National Laboratory (LLNL) electron beam ion traps, EBIT-I
(Run-I) and SuperEBIT (Run-II). The details of their operation have
been described elsewhere \citep{beiersdorfer08a, beiersdorfer03a,
  marrs94a, levine88a, marrs88a}. In brief, highly charged ions are
produced, trapped, and excited by EBIT using a near mono-energetic
electron beam and an electrostatic trap. Several methods have been
developed to inject elements for study
\citep{igbrown86a,schneider1989a,elliot1995a,ullrich1998a,niles2006a,yamada2007a,magee2014a}.
For the experiments described here, neutral sulfur and silicon were
injected into the EBIT's trap region as gaseous $\mathrm{SF}_6$ and
$\mathrm{C}_{10}\mathrm{H}_{30}\mathrm{O}_3\mathrm{Si}_4$,
respectively, using a well collimated ballistic gas injector. Once the
neutral material intersects the electron beam, the molecules are
broken apart and resulting atoms are collisionally ionized and
trapped. To avoid the build up of high-$Z$ material, such as tungsten
and barium originating from the electron gun, the trap region is
emptied and refilled periodically, on a time scale of tens of
milliseconds.

\begin{figure}
\includegraphics[width=0.95\columnwidth]{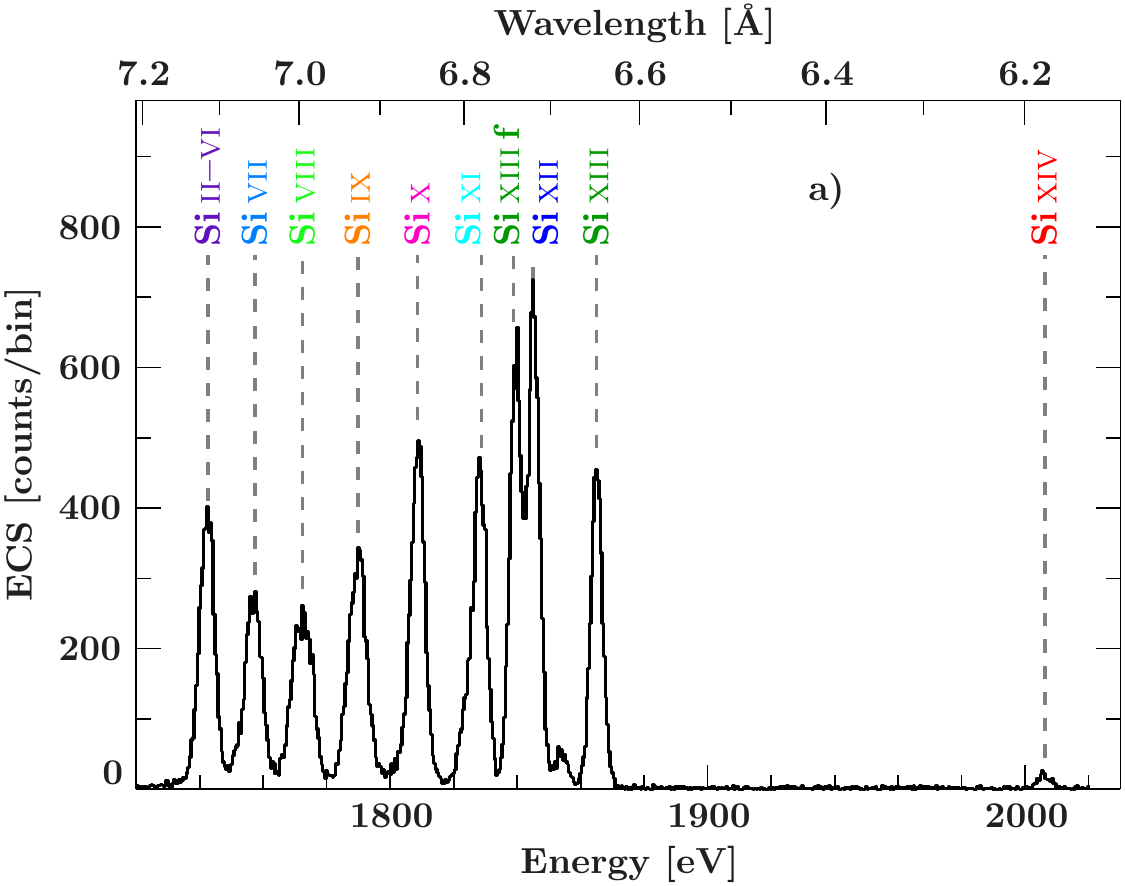}
\includegraphics[width=0.95\columnwidth]{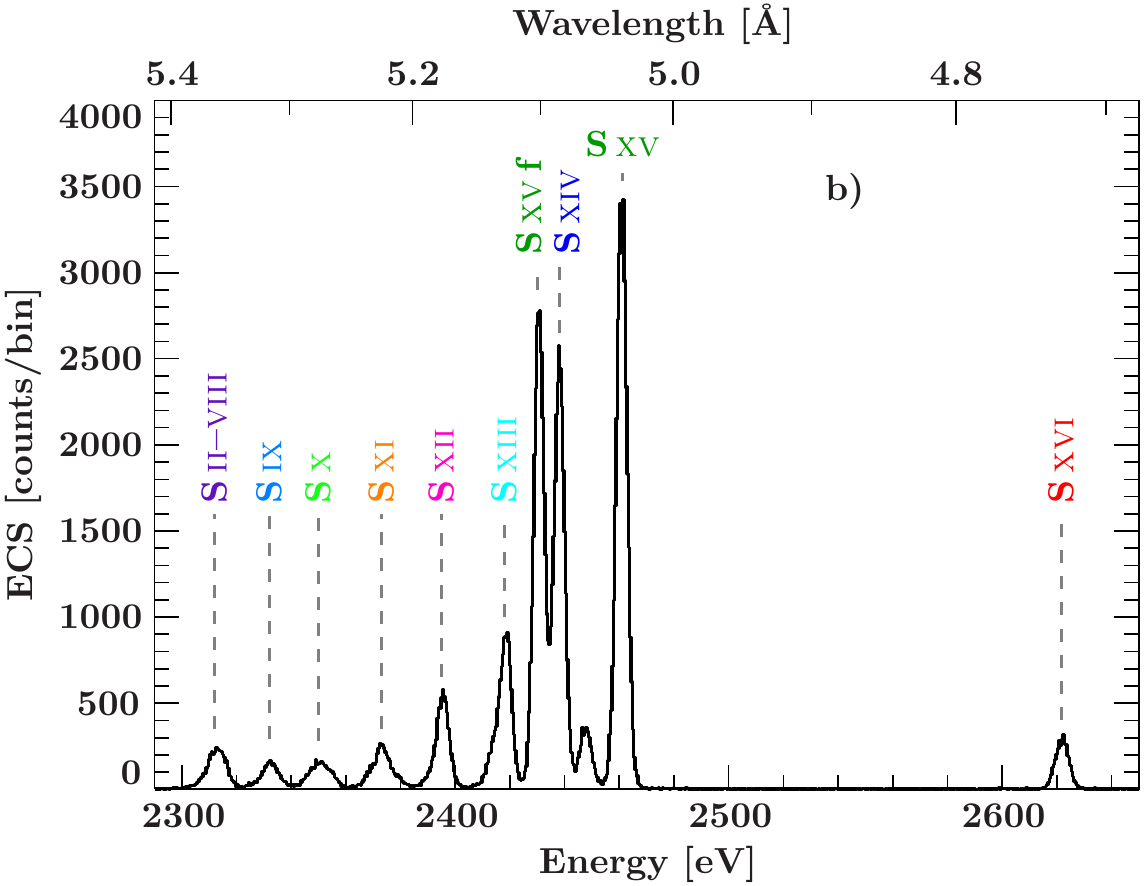}
\caption{Calibrated and summed \textsl{a)} silicon and \textsl{b)} 
   sulfur spectra from all 16 low-energy ECS pixels (Run-I). The color code
   of the ion labels is used whenever we specifically distinguish between
   isoelectronic sequences throughout this work.
   \label{fig:spectra}}
\end{figure}

The electron impact excitation energies of the K-shell transitions in
the silicon and sulfur ions are $\gtrsim$1.73\,keV, while the
ionization energies for the L-shell ions range from 166.8\,eV for
Ne-like Si\,{\sc v} to 707.2\,eV for Li-like S\,{\sc xiv}
\citep{cowan81a}. Hence, in order to excite the K$\alpha$ lines the electron
beam energy must be $\sim3$--10 times the ionization threshold. Under
typical operating conditions at these energies, the charge state
distribution would be dominated by lithium- and helium-like ions. In
order to produce a significant amount of lower charge states at the
high electron impact energies required for inner-shell excitation,
several methods have been developed \citep{decaux1993a,schmidt2004a}.
In the present experiment, the
neutral gas injection pressure is set to values several orders of
magnitudes larger than EBIT's base pressure of
$\lesssim10^{-10}\,\mathrm{Torr}$, short refill cycle times and
relatively low electron beam currents were employed. Together, these
operating parameters yield a significant fraction of low charge states
at high electron impact energy. The spectral signature of significant
amounts of several L-shell ions can easily be seen in the X-ray
spectra (see Figure~\ref{fig:spectra}). Note that the electron beam
energies employed at these measurements were well away from any
dielectronic recombination resonances of the respective measured
elements, i.e., the emission lines originate entirely from electron
impact excitation and inner-shell ionization, contrary to the laser
experiments reported by \citet{faenov94a}.

The spectrum of the X-ray radiation from the trapped ions is recorded
with the 16 low-energy pixels of the EBIT Calorimeter Spectrometer
\citep[ECS;][]{porter08b,porter08c,porter09a,porter09b}, designed and
built by the NASA/GSFC Calorimeter group. The ECS is the improved
successor of the XRS/EBIT \citep{porter04a,porter08a}, the engineering
model of the \textsl{Astro-E2}/\textsl{Suzaku} calorimeter. The energy
resolution of the ECS for these measurements was 4.5--5.0\,eV, typical
for the ECS.  The spectra shown here are similar in quality to a
spectrum measured with the Soft X-ray Spectrometer (SXS) system
\citep{mitsuda10a} aboard the \textsl{Astro-H}/\textsl{Hitomi} X-ray
observatory \citep{takahashi10a} or in the planned X-IFU instrument on
\textsl{Athena} \citep{athena,ravera2014a}.

To assess the systematic errors in our measurement, we conducted a
second experimental run using SuperEBIT (Run-II). SuperEBIT is the
high-energy variant of EBIT-I used in Run-I and can achieve electron beam
energies up to 250\,keV \citep{beiersdorfer03b}. SuperEBIT was used
for Run-II because of beam time availability.

\subsection{Calibration}\label{sec:calib}

Because of slight variations in performance, each pixel in the ECS
array is calibrated separately. The energy scale for each pixel is
determined by fitting $4^\mathrm{th}$ order polynomial functions to
the measured pulse heights in volts space of known reference emission
lines \citep{porter1997a,cottam05a}; here, the X-ray line emission
from K-shell transitions in He-like ions (K$\alpha$ / line w:
$\mathrm{1s\,2p\,}{}^1\mathrm{P}_1\rightarrow1\mathrm{s}^2\,{}^1\mathrm{S}_0$;
K$\beta$: $\mathrm{1s\,3p}\rightarrow1\mathrm{s}^2$; K$\gamma$:
$\mathrm{1s\,4p}\rightarrow1\mathrm{s}^2$) and H-like ions
(Ly\,$\alpha$: $\mathrm{2p}\rightarrow\mathrm{1s}$; Ly\,$\beta$:
$\mathrm{3p}\rightarrow\mathrm{1s}$). Specifically, for the Run-I
measurement, the 1.7 to 1.9\,keV band containing the lower charge
states of silicon was calibrated with K$\alpha$, Ly\,$\alpha$, K$\beta$,
and Ly\,$\beta$ lines of neon and silicon. For the 2.3 to 2.5\,keV band
containing the lower charge states of sulfur, K$\alpha$, Ly\,$\alpha$,
and K$\beta$ of sulfur and K$\alpha$--K$\gamma$ of fluorine were used.
For Run-II, Ne and S K$\alpha$, Ly\,$\alpha$, and K$\beta$, and Si
K$\alpha$--K$\gamma$ and Ly\,$\alpha$ were used to calibrate the silicon
spectra, and Ne K$\alpha$, Ly\,$\alpha$, and K$\beta$, Si and S
K$\alpha$ and Ly\,$\alpha$, and Ar K$\alpha$ were used to calibrate the
sulfur spectra.

The reference wavelengths of the He-like systems used for calibration
originate from \citet{drake88a} in case of the
$\mathrm{1s\,2p}\rightarrow1\mathrm{s}^2$ resonance line labelled ``w''
in the notation of \citet{gabriel72a}.  The wavelengths for
$\mathrm{1s\,3p}\rightarrow1\mathrm{s}^2$ K$\beta$ and
$\mathrm{1s\,4p}\rightarrow1\mathrm{s}^2$ K$\gamma$ Rydberg states
were taken from \citet{vainshtein85a} and corrected for the ground
state of \citet{drake88a} according to \citet{beiersdorfer89a}. Values
for the Lyman series in the H-like systems are from \citet{garcia65a}.
The wavelengths were converted to energy using $E=hc\lambda^{-1}$
where $hc=12398.42\,\mathrm{eV\,\mbox{\AA}}$ \citep[with values for $h$,
$c$ and $e$ from CODATA 2014,][]{codata2014}.

\subsection{Quality of the Calibration\label{sec:qual}}

After calibration, the ECS events were binned to an energy grid of 0.5\,eV.
Figure~\ref{fig:spectra} shows the summed Si and S spectra of all 16
low-energy ECS pixels for Run-I. 
To gauge the accuracy of the energy scale, the location of the H-like
Ly\,$\alpha$ lines and the He-like line w of Si and S are determined
from a simultaneous fit of the calibrated Run-I and Run-II spectra.
The fitted values are then compared to the initial reference values.
Table~\ref{tab:reflines} shows the value from the comparison as well
as from our FAC calculation, which is used as a guide for line
identification (see below, Section~\ref{sec:lineIDFAC}). For silicon line w,
the calibrated values are 0.16\,eV lower than the reference values,
for sulfur line w, they are 0.017\,eV lower. For the S Ly\,$\alpha$ lines, the
difference between theory and experiment is slightly larger, but still
well below 0.5\,eV (Table~\ref{tab:reflines}). Combining the
uncertainties of the Ly\,$\alpha$ and w lines amounts to 0.13\,eV for silicon
and 0.23\,eV for sulfur, which are taken as the systematic
uncertainties. FAC results agree with \citet{drake88a} to within
0.2\,eV in case of the transition energies in He-like ions, and within
0.03\,eV for the transition energies in H-like ions.

The fitted widths of the He-like lines of about 4.5--5.0\,eV are
consistent with the expected energy resolution of the ECS in this
energy region.

\begin{deluxetable*}{llllllll}
\tablecaption{Calibration results\label{tab:reflines}}
\tablecolumns{8}
\tablewidth{0pt}
\tablehead{
\colhead{$Z$} & \colhead{line} & \colhead{FWHM} &
 \multicolumn{3}{c}{line energy (eV)} & \colhead{$\Delta
 E_\mathrm{ref}$} & \colhead{$\Delta E_\mathrm{FAC}$}\\
 \cline{4-6} 
 \colhead{} & \colhead{} & \colhead{(eV)} & \colhead{fit} &
 \colhead{reference} & \colhead{FAC} & \colhead{} & \colhead{}}
\startdata
Si & w & $4.36^{+0.08}_{-0.12}$/$4.92\pm0.12$ & $1864.84\pm0.05$ & 1864.9995  & 1864.812 & $-0.16$ & $0.03$\\
Si & Ly\,$\alpha$ & --- & $2005.59^{+0.17}_{-0.20}$ & 2005.494$^a$ & 2005.516$^a$ & $\phantom{-}0.10$ & 0.07\\
S & w & $4.55\pm0.04$/$4.98\pm0.14$ & $2460.609\pm0.018$ & 2460.6255 & 2460.417 & $-0.017$ & 0.191\\
S & Ly\,$\alpha_1$ & --- & $2622.97^{+0.18}_{-0.26}$ & 2622.700 & 2622.730 & $\phantom{-}0.27$ & 0.24\\
S & Ly\,$\alpha_2$ & --- & $2620.00^{+0.21}_{-0.34}$ & 2619.701 & 2619.731 & $\phantom{-}0.30$  & 0.27\\
\enddata

\tablenotetext{a}{Mean value of Ly\,$\alpha_1$ and Ly\,$\alpha_2$
weighted by their statistical weights.}

\tablecomments{Comparison between the fitted line centers (fit)
of the He-like $\mathrm{1s\,2p}\rightarrow1\mathrm{s}^2$ line w with
the reference value (reference) of \citet{drake88a} and of the H-like
$\mathrm{2p}\rightarrow\mathrm{1s}$ Ly\,$\alpha$ lines of Si and S with the
values of \citet{garcia65a}, which were used for calibration. The full
width half maximum (FWHM) determined from line w (used as detector
resolution throughout the fits) is listed for Run-I / Run-II. $\Delta
E_i$ gives the difference between the fit and the respective
theoretical values. Listed uncertainties are purely statistical.}
\end{deluxetable*}

\section{Spectral Analysis}\label{sec:highresanalysis}
\subsection{Fit Method}\label{sec:fitmethod}

\begin{figure}
 \includegraphics[width=\columnwidth]{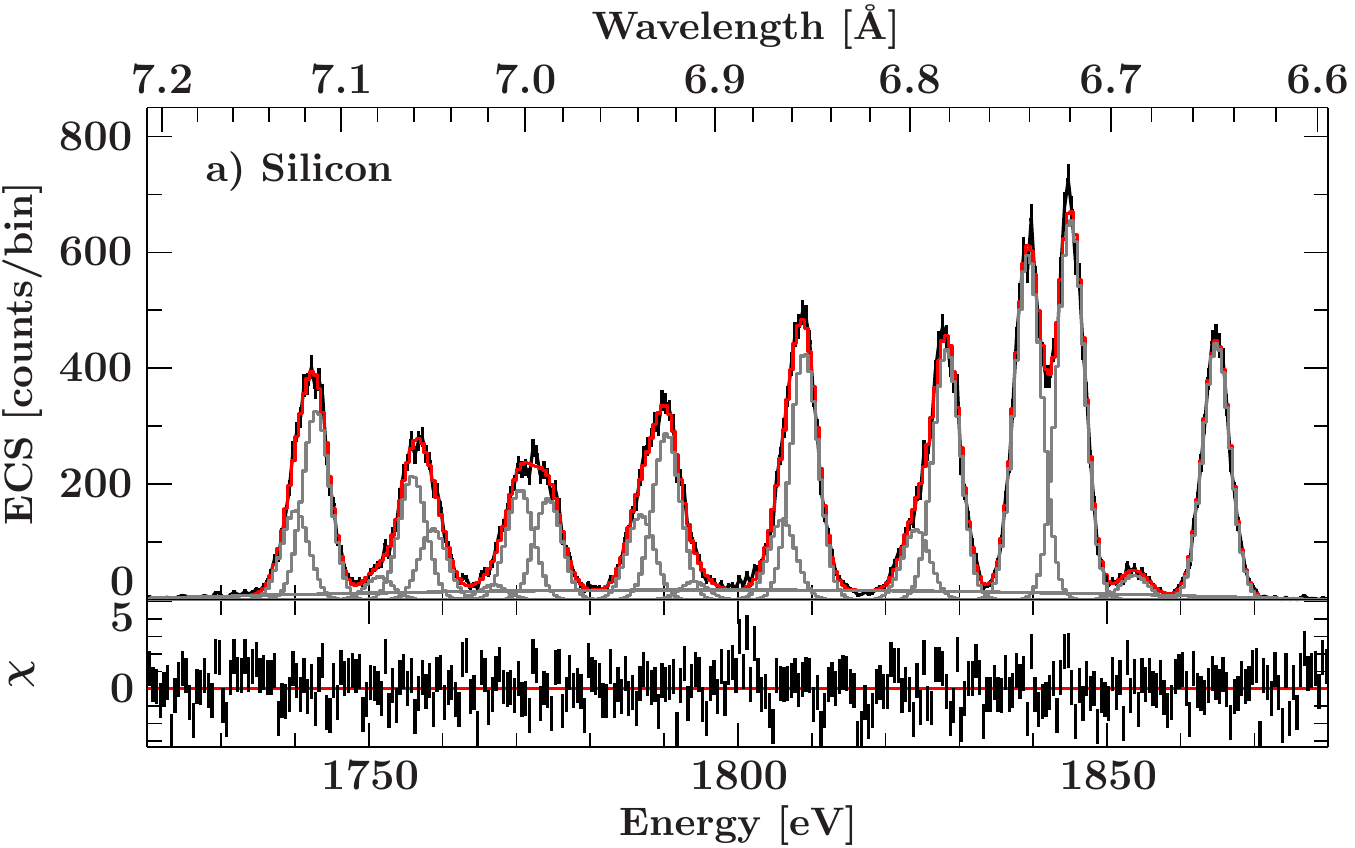}\\
  \includegraphics[width=\columnwidth]{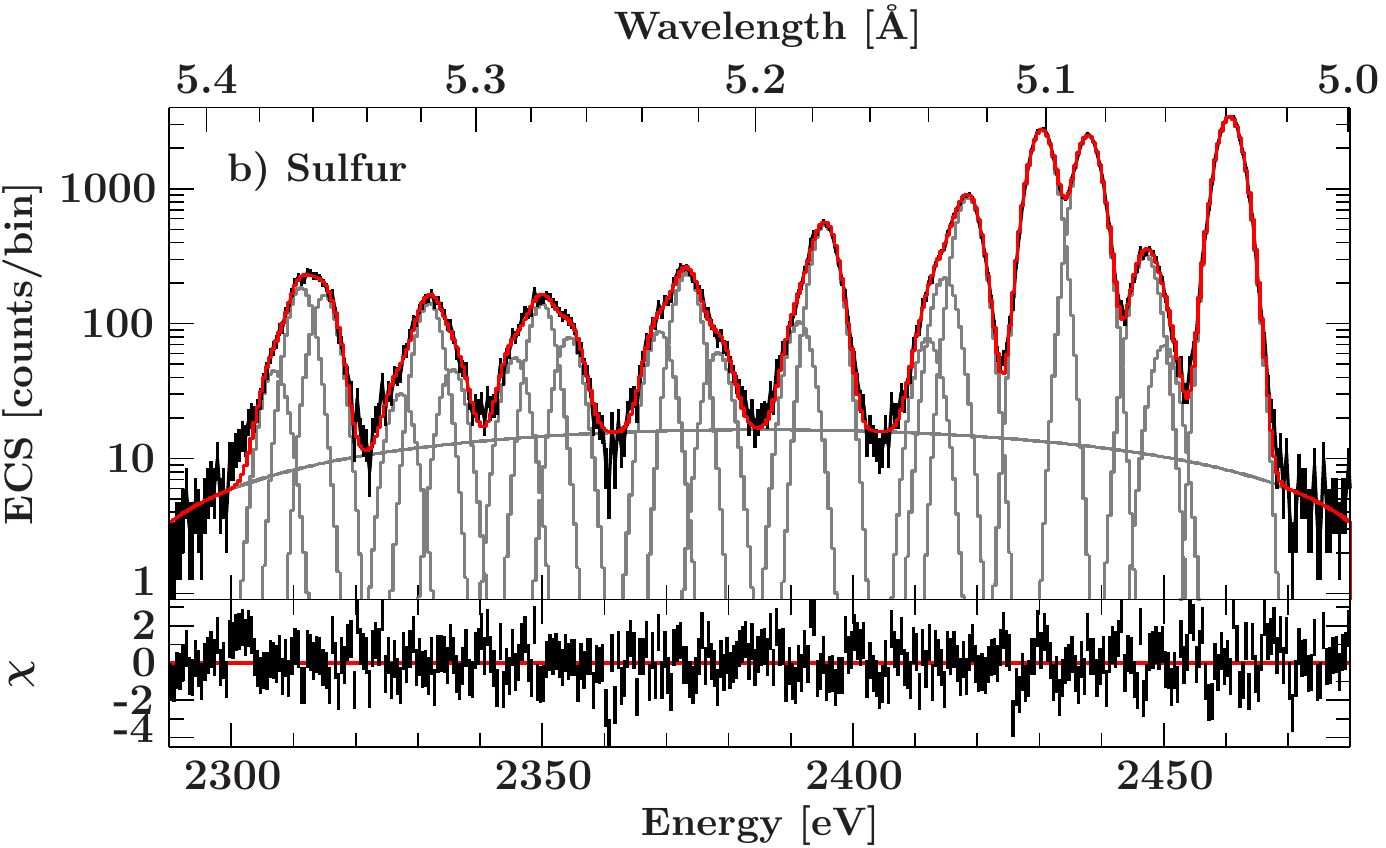}

\caption{Overview over the components fitted to the Si (top) and S
 (bottom) spectra. The
   data are shown in black, the red line shows the total model,
   model components are gray.}   
 \label{fig:sicomp}
  \label{fig:scomp}
\end{figure}

In order to determine the transition energies of as many individual
lines as the data allow, the spectra from Run-I and Run-II were fitted
simultaneously for each element, using the Interactive Spectral
Interpretation System ISIS \citep{houck00a,houck02a,noble08a}. The
modeled energy range spans 1720--1880\,eV for the Si spectra and
2290--2480\,eV for the S spectra. The models for Run-I and Run-II
consist of a sum of individual Gaussian lines, where the centers of
these lines are tied between Run-I and Run-II, their widths are fixed
to the respective resolution (Table \ref{tab:reflines}), and their
normalizations are left to vary freely. Fixing the line widths is
valid because the natural line widths and the Doppler widths are small
compared to the resolution of the calorimeter, no other line
broadening mechanism is present in these experiments, and the energy
resolution of the calorimeter is constant over these small energy
ranges.
In order to account for the flux above background found between the main peaks of
the spectra, e.g., Figure~\ref{fig:scomp}, the models include a single second
order polynomial for each dataset. 
A possible explanation for the presence of this continuum are weak
unresolved lines (see Figure~\ref{fig:palmeri-fac} in
Section~\ref{sec:facvspalm}), low-energy spectral redistribution due
to photon and electron escape events \citep{cottam05a}, or some
combination of both.

In order to determine the number of Gaussian components required to
describe the data, we test the statistical significance of each line.
A Monte Carlo type simulation (see \texttt{mc\_sig} of the Remeis
ISISscripts), generates $10^3$ realizations of fake spectra based on
the existing best fit model: for each energy bin of the fake spectrum,
it draws a random number from a Poisson distribution with the mean
equal to the modeled value. These fake spectra are fitted with the
model used to create them (model~A) and with a model containing an
additional Gaussian line (model~B). Because of the increased number of
degrees of freedom, the $\chi^2$ value for model~B will be at least
slightly better than the $\chi^2$ for model~A. The additional line in
model~B is only accepted if the improvements,
$\Delta\chi^2_{\mathrm{fake},i} = \chi^2_{\mathrm{B},i} -
\chi^2_{\mathrm{A},i}$, of 99\% of the simulated cases are smaller
than the improvement in the real spectrum. Figure~\ref{fig:sicomp}
shows the final distribution of the single Gaussian components for
silicon and for sulfur. Tables~\ref{tab:sifit} and \ref{tab:sfit} list
the resulting line centers with their statistical 90\% confidence
limits.

As an additional consistency check for the accuracy of our results, in
a second approach we allow for a constant shift of the Run-II data compared
to the Run-I data. The derived constants of $0.13^{+0.06}_{-0.05}$\,eV
for Si and $-0.12\pm0.05$\,eV for S are consistent with our estimate
of the systematic uncertainty of our calibration (Section~\ref{sec:qual}).

\subsection{Line Identification with FAC}\label{sec:lineIDFAC}

To identify the lines associated with our measured spectra, we use FAC
\citep{gu04b,gu08a} to calculate the wavelengths of transitions in the
involved ions and model the
measured spectra. FAC is a compound package based on a fully
relativistic ansatz via the Dirac equation which provides functions to
calculate the atomic structure, bound-bound and bound-free processes,
and includes a collisional ionization equilibrium code to estimate the
line intensities for given plasma conditions \citep[electron beam
  properties or plasma temperatures;][]{guFAC}. The accuracy of FAC,
determined from comparisons between FAC and experiments, is a few eV
or 10--30\,m\AA\ at $\sim$10\AA\ for energy levels (other than
H-like) and 10--20\,\% for radiative transition rates and cross
sections \citep{guFAC}.

Our FAC calculations take into account radiative
\mbox{(de-)}excitation, collisional \mbox{(de-)}excitation and
ionization, autoionization, dielectronic recombination, and radiative
recombination. At EBIT densities, the coronal limit applies, i.e.,
electron impact collisional excitation, inner-shell ionization, and
subsequent radiative cascades are the main processes to populate upper
states. At the electron beam energies used here, no emission from
dielectronic recombination exists for the ions of interest and no
X-rays from radiative recombination fall into our energy band.
Although the main application for our results is photoionized plasmas,
the collisional nature of EBIT does not compromise this task.

Our calculations include emission from all the $n\rightarrow1$
transitions in Na-like to H-like silicon and sulfur, where $2\leq
n\leq 5$, allowing interactions between all levels, including $\Delta
n=0$ transitions. For these limits, the calculation could be completed
in a reasonable time. The contribution to the line strength from
higher $n$ transitions is negligible. Since the charge state
distribution in EBIT depends on ionization and recombination
processes, the level populations are estimated for all ions in a
single calculation. The other plasma code parameters are the electron
beam energy, which we assume to follow a Gaussian distribution with an
energy spread of $\sim40$\,eV \citep{beiersdorfer1992a,gu99a}, and an
electron density of $10^{12}\,\mathrm{cm}^{-3}$, which we estimate
from beam current and energy. The relative abundances of the trapped
ions are set to be 1. The simulation of the spectrum produced in the
trap is therefore not self-consistent.

\begin{figure}
\includegraphics[width=\columnwidth]{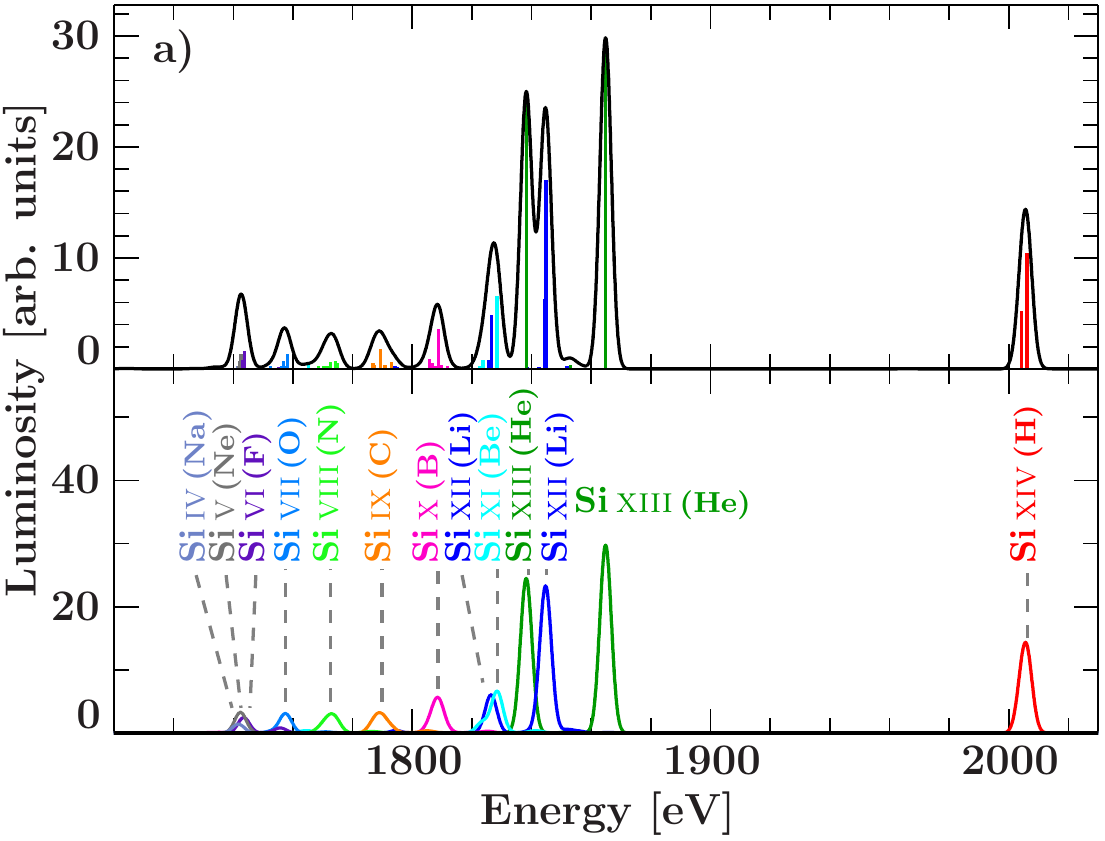}
\includegraphics[width=\columnwidth]{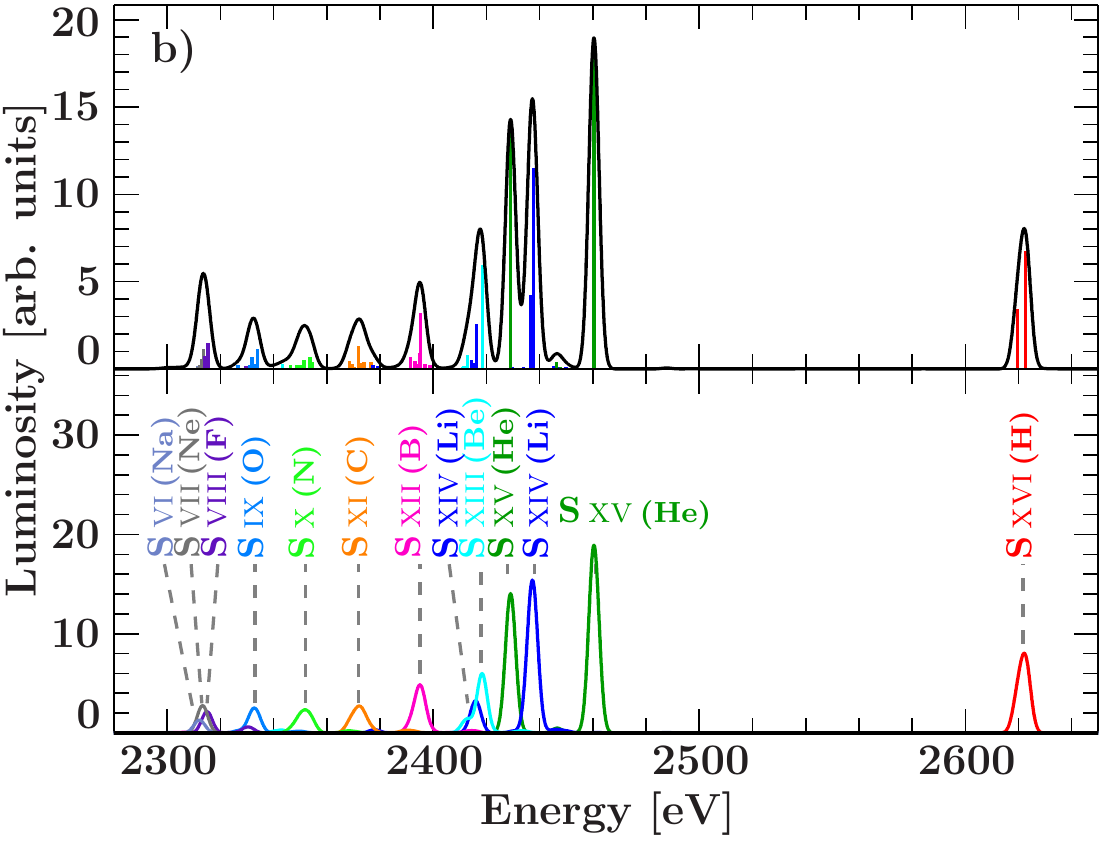} 
\caption{FAC simulation of the \textsl{a)} Si and \textsl{b)} S K$\alpha$
  spectra. For each subfigure the top panel shows the transition
  energies with their predicted luminosity and the total spectrum
  (black line) resulting from a convolution with Gaussians at the
  resolution of the ECS. The bottom panels show the convoluted spectra
  individually for each ionization state, which sum to
  the black line of the top panel. The impact of line blends can be
  seen quite clearly. Labels include the
  corresponding iso-electronic sequence in parentheses.}\label{fig:facsi}
\end{figure}

Figure~\ref{fig:facsi} illustrates the resulting FAC simulations for
silicon and sulfur, considering the presence of H- through Na-like
ions. The line centers of transitions calculated by FAC are convolved
with a Gaussian line with a FWHM of 4.6\,eV, i.e., the resolution of
the calorimeter (see Section~\ref{sec:qual}).

While the strongest K-shell line features from each charge state are
easily resolved (Figures~\ref{fig:spectra} and~\ref{fig:facsi}),
identifying the transitions that contribute to each feature is
accomplished by comparison to the FAC calculation as follows. For each
feature we plot the data and individual model components and overlay
them with the transitions obtained from FAC (Figures~\ref{fig:si-fits}
and~\ref{fig:s-fits}). Then we assume that for every Gaussian fit
component the main contribution comes from the strongest FAC lines at
this energy and identify the model component with these lines. The
results are listed in Tables~\ref{tab:sifit} (Si) and~\ref{tab:sfit}
(S). In each row the FAC lines are followed by the corresponding
transitions as calculated by P08 (see Section~\ref{sec:facvspalm} for
details) and CHIANTI, if available.
For most measured peaks, the distribution of the FAC lines agrees well
enough with the measurements to allow a reliable identification.  Both
the Si and S spectra behave very similarly, so our description of the
spectra here focuses on the contributions by iso-electronic sequence,
for the most part not distinguishing in $Z$ except in the rare cases
where significant differences occur between the Si and S spectra.

The main Li-like, Be-like, and B-like features are each dominated by a
single strong transition that is easily reproduced by the Gaussian
components fitted to the spectra (see features labeled Li-2, Be-1, B-1
in Figures~\ref{fig:si-fits} and~\ref{fig:s-fits} panels e-g).
Although there are a few weaker transitions surrounding these strong
lines, they do not strongly affect the fitted line centers. Both the
Be- and B-like features have a low-energy shoulder caused by weaker
transitions that have a just large enough separation from the strong
transition to be resolved. According to our FAC calculations, the
Li-like ion also has a relatively strong transition that sits right
between the Be-like lines. Although in the synthetic Si spectrum the
Li-like line appears to have similar strength as the strong Be-like
line, a comparison of Figure~\ref{fig:facsi} to the measured Si
spectrum shows that due to the incorrect assumption of charge balance
entering our simulation, the synthetic spectra overestimate the
Li-like features relative to the Be-like ones. Accordingly, despite
this Li-like transition being unresolved in Si, it does not seem to
affect the fitted line centers of the Be-like transitions Be-1 and
Be-2 much (Figure~\ref{fig:si-fits}, panel f). For S on the other hand
(Figure~\ref{fig:s-fits}, panel f), the Li-like transitions are
attributed to their own Gaussian component (Be-2) while the weak
Be-like line is assigned to a separate component (Be-3).

The transition rich spectra of the lower charge states C-like, O-like,
and N-like are more complex as they have many transitions of similar
strength rather than a distinct strong transition among a few weak
ones (Figures~\ref{fig:si-fits} and~\ref{fig:s-fits}, panels b-d).
However, some of these transitions tend to cluster into groups. The
separation of these groups is larger for the higher-$Z$ element S,
making it easier to partially resolve them. As discussed for iron by
\citet{decaux97a}, starting around C-like ions the K$\alpha$ line
emission of the lower charge states probably has strong contributions
from states excited through inner-shell ionization in addition to the
collisionally excited states.

In the C-like ions (Figures~\ref{fig:si-fits} and~\ref{fig:s-fits},
panel d), the strongest fitted component, C-2, is made up of
the strongest calculated transitions at slightly lower energies than
the component's center and a few weaker transitions at and slightly
above the fitted energy. The C-like feature also has a strong
low-energy shoulder (C-3) from transitions similar in strength to the
ones from the C-2 cluster, and a weaker high-energy shoulder (C-1)
consisting of a C-like and two weak Li-like transitions.

The N-like transitions split into four main groups
(Figures~\ref{fig:si-fits} and~\ref{fig:s-fits}, panel c). They are
accompanied by a Be-like transition in the low-energy tail of their
spectral feature. Again, the larger spacing in S is beneficial,
although in both Si and S this feature is modeled by three components.
While N-1 coincides well with the first group of calculated
transitions on the high-energy side for both components, the second
group containing the other two of the strongest four transitions falls
right between N-1 and N-2 in Si, but is clearly attributed to N-2 in
S. N-2 also encompasses the third group of transitions, while N-3
contains the last group of N-like transitions and the mentioned
Be-like transition.

The O-like peak is also described by three Gaussians
(Figures~\ref{fig:si-fits} and~\ref{fig:s-fits}, panel b). The
strongest line calculated with FAC makes up the weaker component at
high energies (O-1), while the main component (O-2) consists of a
number of weaker transitions. A single O-like transition accounts for
the low-energy shoulder (O-3).

\begin{figure}
\includegraphics[width=\columnwidth]{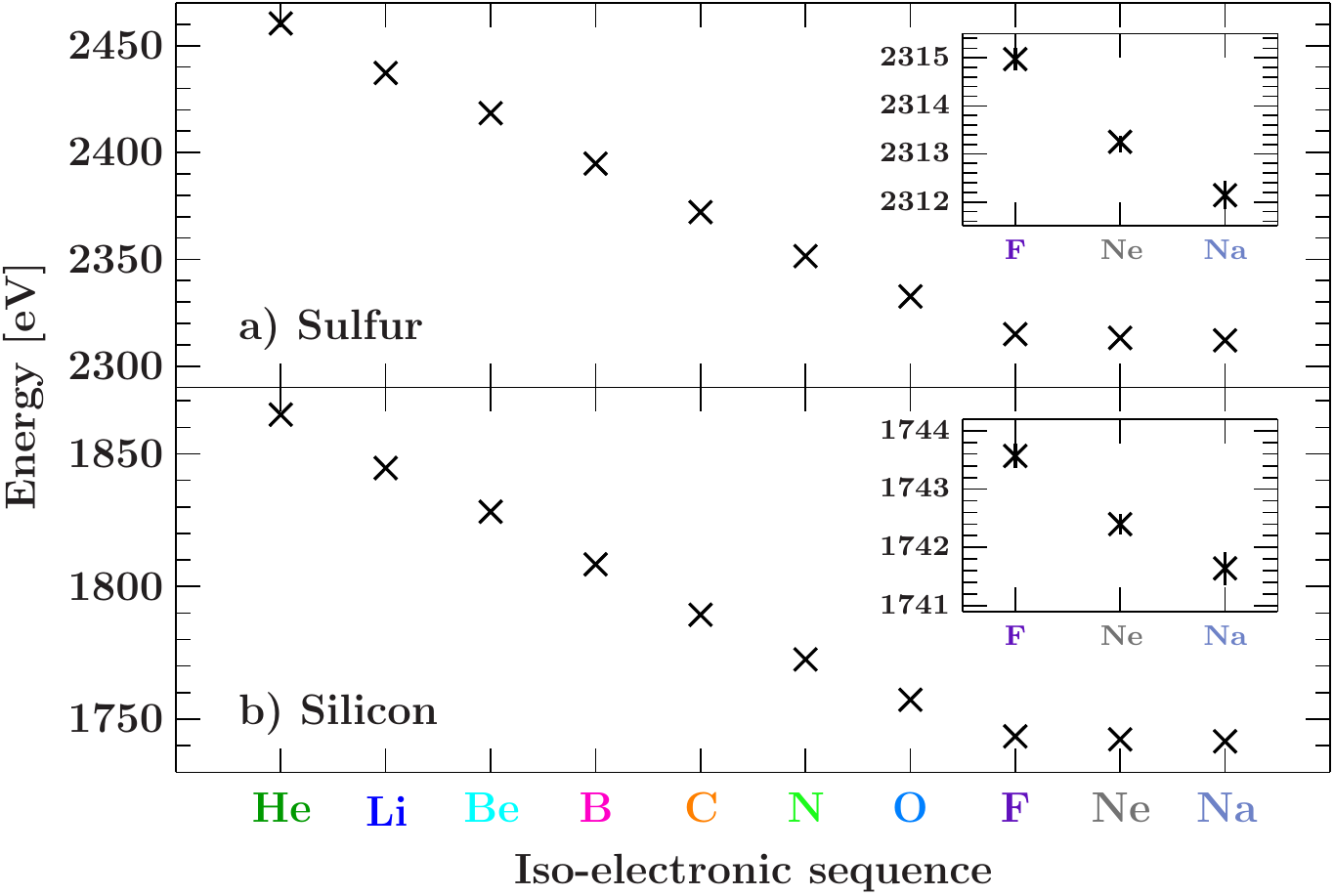}

\caption{Energies of the line centers for different ionization stages
  of sulfur (a) and silicon (b), derived from FAC calculations. Lines
  for ions with more than 9 electrons (F-like) blend strongly with a
  predicted energy spacing of $\sim1$\,eV between charge states.}
\label{fig:cevvsion}
\end{figure}

\begin{figure}
\includegraphics[width=\columnwidth]{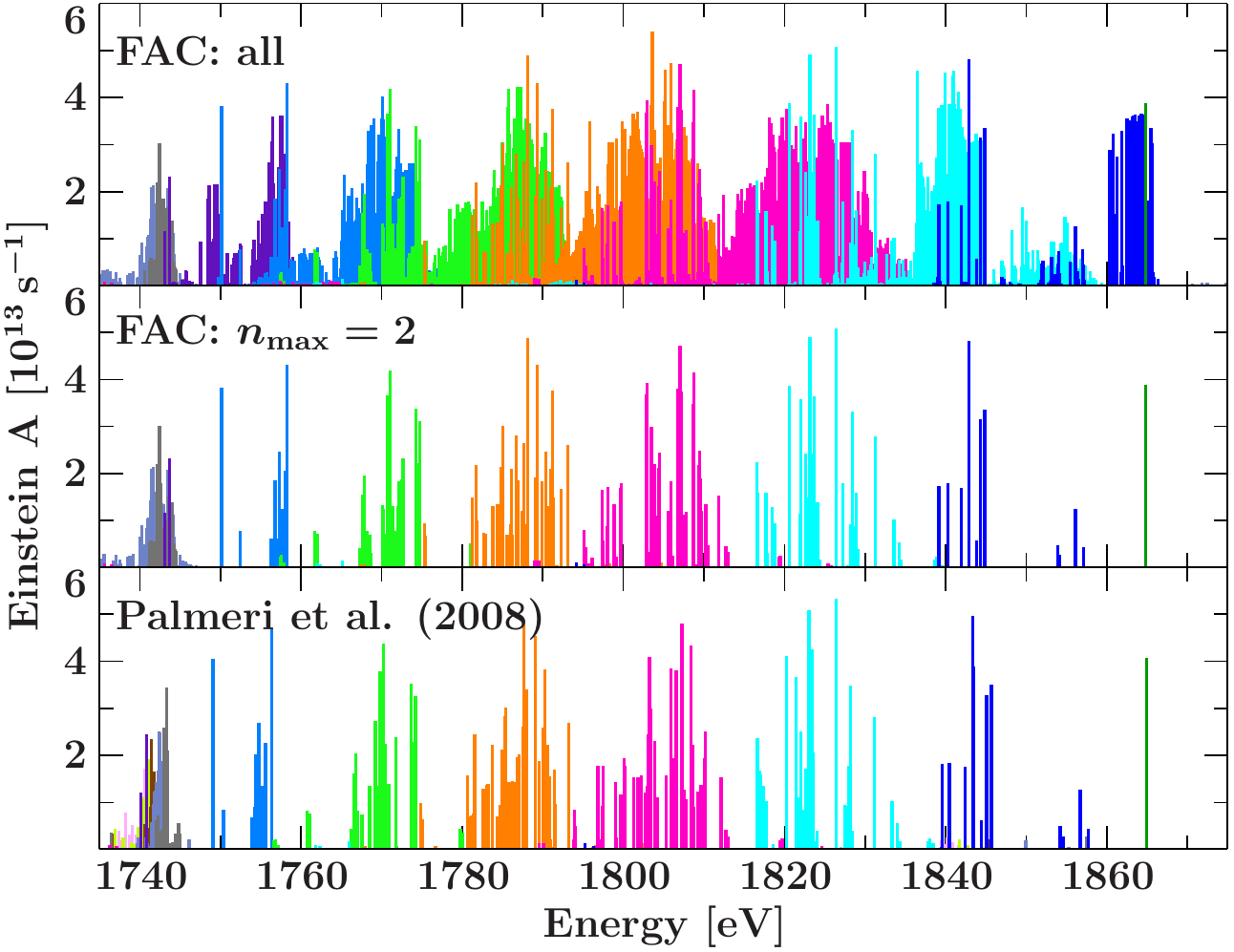}
\caption{Comparison between the atomic data of Si presented by
  P08 (\textsl{bottom}) and the calculation with FAC (\textsl{middle}: only
  $2\ell\rightarrow1\mathrm{s}$ transitions as in P08; \textsl{top}:
  all calculated transitions in this shown energy range, including
  satellites with an electron in up to the $n_\mathrm{max}=5$ shell)
  for He- through Ne-like ions. Different colors represent
  different ionization states (see Figure~\ref{fig:spectra}).}
 \label{fig:palmeri-fac}
\end{figure}

The lowest energy peak (Figures~\ref{fig:si-fits}
and~\ref{fig:s-fits}, panel a) consists of a blend of K-shell
transitions in F-like ions as well as emission from lower charge
states (Figure~\ref{fig:cevvsion}). This is a result of the fact that,
for charge states other than F-like, emission is dominated by
innershell ionization followed by radiative decay in these cases and
the effect of additional spectator electrons in $n\geq3$ shell on
these transition energies is relatively small. Additionally, owing to
the open $n=3$ shell, the M-shell ions have a more complex energy
level structure -- and, therefore, a multidue of transitions -- in
each of these charge states. The energy ranges covered by these
transitions overlap severely (Figure~\ref{fig:palmeri-fac}).
Specifically, the K$\alpha$ transition energies from these charge
states fall within a 3\,eV energy band and are therefore unresolved
(Figure~\ref{fig:cevvsion}). Consequently, although the F-like ion
only has two distinct transitions, we cannot resolve this charge state
individually from the transitions in M-shell ions in these low-Z
elements. This last peak is modeled by two (Si) and three (S) Gaussian
components, respectively. In both cases, we attribute the first, i.e.,
high-energy component (F-1) to a mixture of transissions in F-like and
Ne-like ions. In case of the \ion{Si}{10}-1 line at 1740.04\,eV,
however, there are no lines of considerable strength in our
calculations that could be used for identification. We tentatively
identify this line as a blend of K$\alpha$ emission from very low
charge states with more than 10 electrons. Similarly, although
Table~\ref{tab:sfit} lists weak transitions in Na-like \ion{S}{6} and
B-like \ion{S}{12} for the lines F-2 and F-3, these lines probably
also have a significant contribution from  weak lines
from near-neutral ions, as discussed for the case of silicon.

Also notable is that for both Si and S, line z as calculated with FAC
(Si: 1838.20\,eV, S: 2429.075\,eV) has a large offset ($>$1\,eV)
compared to the measured line center (Si: 1839.33\,eV, S:
2430.380\,eV). Our measurement is, however, consistent with the
reference values reported by \citet[Si: 1839.448\,eV; S:
  2430.347\,eV]{drake88a}.

\subsection{Comparison with \citet{palmeri08a}}\label{sec:facvspalm}

For completeness and to provide a test for the accuracy of the
K$\alpha$ line energies employed by XSTAR, we compare our measurements
and our FAC calculations to those of P08. Note that the calculated
transition wavelengths listed by P08 have been empirically shifted by
P08 for ions with $3\leq N\leq 9$, where $N$ is the number of
electrons. A qualitative comparison between the results obtained with
FAC and the lines published by P08 is displayed in
Figure~\ref{fig:palmeri-fac} for silicon. Since the P08 data do not
provide luminosities, the line distributions are shown via their
radiative transition rates (Einstein A). P08 only list
$2\ell\rightarrow1\mathrm{s}$ transitions. We therefore also filter
for these lines calculated with FAC. The transitions with a spectator
electron in a higher $n$ shell blend strongly with K$\alpha$ transitions of
the next ionization state, but according to the FAC calculations,
their contribution to the K$\alpha$ line strength is negligible
(Figure~\ref{fig:facsi}).

As expected, the positions of the He-like lines agree very well. For
lower ionization states, the general distribution of the lines is
still similar, but the predicted energy separation of some line
features does not agree. For example, there are two O-like \ion{Si}{7}
lines around 1750\,eV (Figure~\ref{fig:palmeri-fac}), specifically the
transitions
1\lss[2]2\lss2\lsp[5] \sljo{1}{P}{1} -- 1\lss2\lss2\lsp[6] \slj{1}{S}{0} 
 and 
1\lss[2]2\lss[2]2\lsp[4] \slj{1}{S}{0} -- 1\lss2\lss[2]2\lsp[5] \sljo{1}{P}{1} 
(P08) respectively 
$(1\nsf\,2\ns\,2\np[2]\,(2\npp[3])_{3/2})_1$ --
$(1\ns\,2\ns\,2\np[2]\,2\npp[4])_0$ 
and 
1\nsf\,2\nsf\,2\np[2]\,$(2\npp[2])_0$ -- 
$(1\ns\,2\nsf\,2\np[2]\,(2\npp[3])_{3/2})_1$ 
(FAC), for which the ratio of the transition probabilities is
approximately the same in both calculations (P08: 0.21; FAC: 0.20).
The separation of their line energies, however, is almost twice as
large in FAC (2.32\,eV) as in P08 (1.33\,eV). The most outstanding
difference is that in the P08 calculations the two F-like spectral
lines at $\sim1741$\,eV have distinctly lower energies than the
Ne-like lines, although the Ne-like iso-electronic sequence has an
electron more than the F-like ions. This behavior is in contrast to the
FAC calculations where the F-like lines have higher energies.

Comparing FAC ($jj$-coupling) and P08 ($LS$-coupling) is not trivial
since the two calculations are based on different coupling schemes. 
It is therefore necessary to translate one scheme into the other. The
calculations are in sufficient agreement such that most lines can be
identified through a comparison of the line lists instead of resorting
to a complicated formal mapping between both schemes \citep[see,
e.g.,][]{calvert79a,dyall86a}. We do this comparison by first sorting
the levels of both calculations according to energy and then matching
the levels in order of increasing energy. The match is cross-checked
via the total angular momentum $J$, which is the only good quantum
number common between the two coupling schemes and therefore should be
identical between them. In cases where $J$ does not match between two
assigned levels, $LS$-coupling multiplets can be rearranged for their
$J$s to fit the $jj$-coupling partners. This is possible because
within these multiplets the differences between the calculated level
energies are smaller than the estimated uncertainty of the
calculations and, in most cases, smaller than the energy differences
between P08 and FAC results.

\begin{figure}
\vspace*{0.5\baselineskip}

  \includegraphics[width=\columnwidth]{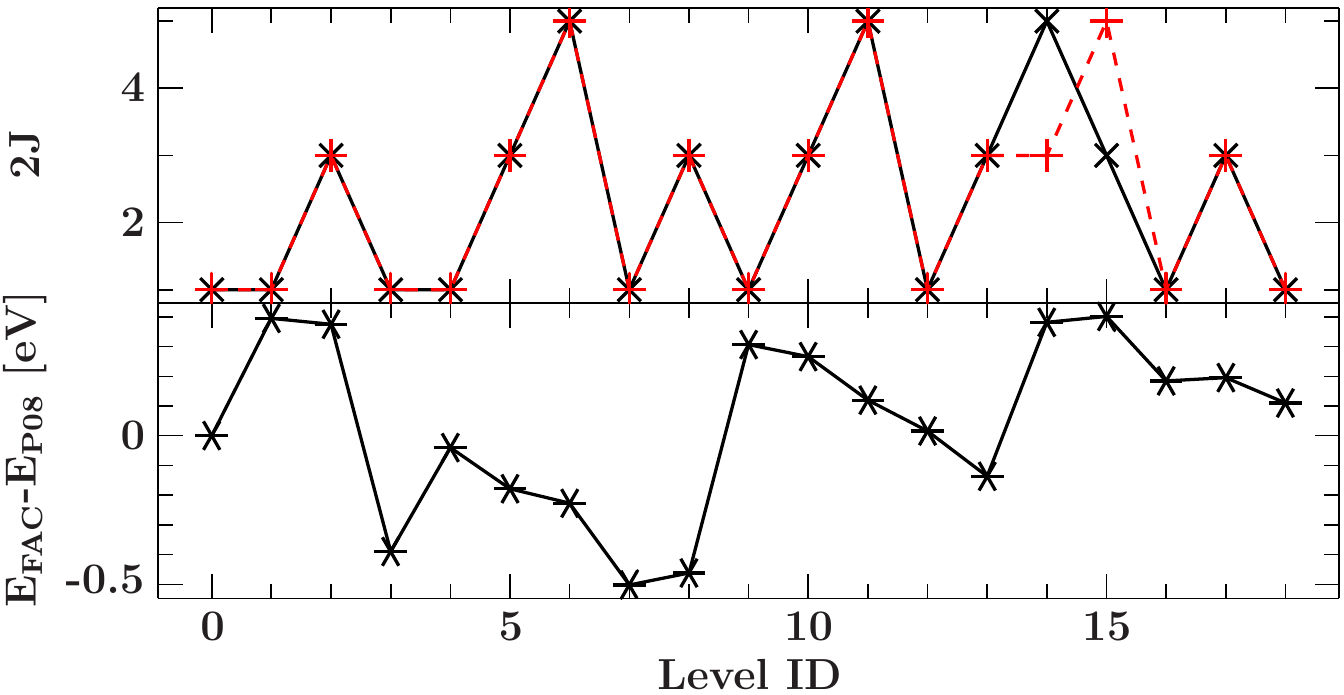}
  \caption{Comparison between FAC and P08 energy levels for Li-like
    Si. The top panel shows two times the total angular momentum
    (black: FAC, red: P08) of the levels sorted by energy, where the
    level ID equal to zero corresponds to the ground state (see
    Table~\ref{tab:matchsili}). The lower panel displays the
    difference between these energies of the two
    calculations.} \label{fig:matchsili}
\end{figure}

As an example, Table~\ref{tab:matchsili} lists the sorted energy
levels of Li-like Si {\sc xii} for both FAC and P08.
Figure~\ref{fig:matchsili} shows a comparison of their total angular
momenta. The energy differences between the levels from different
calculations are within their stated accuracy of at least 2\,eV. The
levels with IDs 14 and 15 are mismatched in their $J$. These levels
belong to the doublet \sle{2}{D}, with a fine structure splitting of
roughly 0.1\,eV. This difference is small compared to the
$\sim$0.4\,eV between FAC and P08. For practical purposes we therefore
assume that FAC level 14 corresponds to P08 level 15 and vice versa.
The results are listed in Tables~\ref{tab:sifit} and~\ref{tab:sfit}.

\begin{deluxetable*}{clcr|lcr}
\tablecaption{Comparison of the FAC and P08 energy levels
   of Li-like Si.\label{tab:matchsili}}
\tablecolumns{7}
\tabletypesize{\scriptsize}
\tablewidth{\columnwidth}
\tablehead{
\colhead{Level} &  \multicolumn{3}{c|}{FAC / $jj$-coupling} & 
 \multicolumn{3}{c}{P08 / $LS$-coupling} \\
 \colhead{ID} & \colhead{label} & \colhead{$2J$} & \colhead{E (eV)}
 & \colhead{label} & \colhead{$2J$} & \colhead{E (eV)}}
\startdata
0 & 1\nsf\,2\ns                              & 1 & 0.0000 & 1\lss[2]2\lss\ \slj{2}{S}{1/2} & 1 &    0.0000 \\
 1 & 1\nsf\,2\np                              & 1 & 24.2019 & 1\lss[2]2\lsp\ \sljo{2}{P}{1/2} & 1 &   23.8072 \\
 2 & 1\nsf\,2\npp                             & 3 & 25.1920 & 1\lss[2]2\lsp\ \sljo{2}{P}{3/2}           & 3 &   24.8172 \\
 3 & 1\ns\,2\nsf                              & 1 & 1819.3742 & 1\lss2\lss[2]\ \slj{2}{S}{1/2}            & 1 & 1819.7636 \\
 4 & $((1\ns\,2\ns)^{}_1\,2\np)^{}_{1/2}$  & 1 & 1825.5977 & 1\lss(\sle{2}{S})\,2\lss2\lsp(\slo{3}{P})\ \sljo{4}{P}{1/2} & 1 & 1825.6379 \\
 5 & $((1\ns\,2\ns)^{}_1\,2\np)^{}_{3/2}$  & 3 & 1825.8735 & 1\lss(\sle{2}{S})\,2\lss2\lsp(\slo{3}{P})\ \sljo{4}{P}{3/2} & 3 & 1826.0523 \\
 6 & $((1\ns\,2\ns)^{}_1\,2\npp)^{}_{5/2}$ & 5 & 1826.5509 & 1\lss(\sle{2}{S})\,2\lss2\lsp(\slo{3}{P})\ \sljo{4}{P}{5/2} & 5 & 1826.7783 \\
 7 & $((1\ns\,2\ns)^{}_0\,2\np)^{}_{1/2}$  & 1 & 1844.2892 & 1\lss(\sle{2}{S})\,2\lss2\lsp(\slo{3}{P})\ \sljo{2}{P}{1/2} & 1 & 1844.7912 \\
 8 & $((1\ns\,2\ns)^{}_0\,2\npp)^{}_{3/2}$ & 3 & 1844.8632 & 1\lss(\sle{2}{S})\,2\lss2\lsp(\slo{3}{P})\ \sljo{2}{P}{3/2} & 3 & 1845.3252 \\
 9 & $(1\ns\,(2\np[2])^{}_0)^{}_{1/2}$      & 1 & 1851.4807 & 1\lss(\sle{2}{S})\,2\lsp[2](\sle{3}{P})\ \slj{4}{P}{1/2}    & 1 & 1851.1741 \\
10 & $((1\ns\,2\np)^{}_0\,2\npp)^{}_{3/2}$ & 3 & 1851.8761 & 1\lss(\sle{2}{S})\,2\lsp[2](\sle{3}{P})\ \slj{4}{P}{3/2}    & 3 & 1851.6101 \\
11 & $(1\ns\,(2\npp[2])^{}_2)^{}_{5/2}$    & 5 & 1852.4077 & 1\lss(\sle{2}{S})\,2\lsp[2](\sle{3}{P})\ \slj{4}{P}{5/2}    & 5 & 1852.2880 \\
12 & $((1\ns\,2\ns)^{}_1\,2\npp)^{}_{1/2}$ & 1 & 1853.9104 & 1\lss(\sle{2}{S})\,2\lss2\lsp(\slo{1}{P})\ \sljo{2}{P}{1/2} & 1 & 1853.8947 \\
13 & $((1\ns\,2\ns)^{}_0\,2\npp)^{}_{3/2}$ & 3 & 1854.0850 & 1\lss(\sle{2}{S})\,2\lss2\lsp(\slo{1}{P})\ \sljo{2}{P}{3/2} & 3 & 1854.2217 \\
\bf
14 & $((1\ns\,2\np)^{}_1\,2\npp)^{}_{5/2}$ & \bf5 & \bf1864.3543 &
1\lss(\sle{2}{S})\,2\lsp[2](\sle{1}{D})\ \slj{2}{D}{3/2}
& \bf3 & \bf1863.9732 \\
\bf15 & $((1\ns\,2\np)^{}_0\,2\npp)^{}_{3/2}$ & \bf3 & \bf1864.4771
& 1\lss(\sle{2}{S})\,2\lsp[2](\sle{1}{D})\ \slj{2}{D}{5/2}
& \bf5 & \bf1864.0761 \\
16 & $((1\ns\,2\np)^{}_1\,2\npp)^{}_{1/2}$ & 1 & 1867.1243 & 1\lss(\sle{2}{S})\,2\lsp[2](\sle{3}{P})\ \slj{2}{P}{1/2}    & 1 & 1866.9400 \\
17 & $(1\ns\,(2\npp[2])^{}_2)^{}_{3/2}$    & 3 & 1868.0803 & 1\lss(\sle{2}{S})\,2\lsp[2](\sle{3}{P})\ \slj{2}{P}{3/2}    & 3 & 1867.8849 \\
18 & $(1\ns\,(2\npp[2])^{}_0)^{}_{1/2}$    & 1 & 1881.3194 & 1\lss(\sle{2}{S})\,2\lsp[2](\sle{1}{S})\ \slj{2}{S}{1/2}    & 1 & 1881.2095 \\
\enddata

\vspace{-4mm}
\tablecomments{
   With the exception of levels 14 and 15 (marked bold)
   which have to be swapped in either FAC or in P08 in order for the
   total angular momentum J to match, this table can be used to match
   the $LS$- and $jj$-coupling notations. The level IDs are the same
   as in Figure~\ref{fig:matchsili}.}
   
\end{deluxetable*}

\begin{deluxetable*}{lllll|lllll}
\tablecaption{Center [eV] of unresolved line blends for Si and S
\label{tab:center}}
\tabletypesize{\footnotesize}
\tablewidth{0pt}
\tablehead{
\multicolumn{5}{c}{Silicon} &
\multicolumn{5}{c}{Sulfur} \\
  \colhead{Ion\tablenotemark{a}} & \colhead{this work} & \colhead{FAC} &
  \colhead{Behar} & \colhead{House} &
\colhead{Ion\tablenotemark{a}} & \colhead{this work} & \colhead{FAC} &
\colhead{Behar} & \colhead{House}}
\startdata
\ion{Si}{12} (Li) & $1845.02\pm0.07$ & $1844.67\pm0.07$ & 1845.83 & 1836.80 &
\ion{S}{14} (Li) & $2437.761\pm0.027$ & $2437.22\pm0.10$ & 2438.71 & 2428.21 \\
 &                  &                  & 1845.28 & &
             &                    &                  & 2437.75 &         \\
\ion{Si}{11} (Be) & $1827.51\pm0.06$\tablenotemark{b} & $1828.20\pm0.18$ & 1829.21 & 1819.82 &
\ion{S}{13} (Be) & $2417.51\pm0.05$\tablenotemark{b}  & $2418.29\pm0.23$ & 2418.73 & 2408.40 \\
\ion{Si}{10} (B)  & $1808.39\pm0.05$ & $1808.38\pm0.16$ & 1808.93 & 1801.57 &
\ion{S}{12} (B)  & $2394.95\pm0.05$   & $2394.78\pm0.18$ & 2395.37 & 2386.61 \\ 
&                  &                  & 1806.30 & &
            &                    &                  & 2392.13 &         \\ 
\ion{Si}{9} (C)  & $1789.57\pm0.07$ & $1789.39\pm0.22$ & 1786.77 & 1784.72 &
\ion{S}{11} (C)  & $2372.81\pm0.09$   & $2372.12\pm0.26$ & 2368.82 & 2366.56 \\ 
&                  &                  & 1790.90 & &
            &                    &                  & 2374.27 &         \\ 
\ion{Si}{8} (N)  & $1772.01\pm0.09$ & $1772.55\pm0.22$ & 1771.46 & 1769.43 &
\ion{S}{10} (N)  & $2350.40\pm0.12$   & $2351.48\pm0.27$ & 2349.97 & 2347.74 \\ 
\ion{Si}{7} (O)  & $1756.68\pm0.08$ & $1757.29\pm0.21$ & 1755.40 & 1755.40 &
\ion{S}{9} (O)  & $2332.06\pm0.10$   & $2332.65\pm0.25$ & 2330.53 & 2330.53 \\
\ion{Si}{6} (F)  & $1742.03\pm0.06$\tablenotemark{c} & $1743.57\pm0.22$ & 1741.60 & 1743.31 &
\ion{S}{8} (F)  & $2312.37\pm0.09$\tablenotemark{c}   & $2314.97\pm0.24$ &
2313.57 & 2314.87 \\
\enddata

\tablenotetext{a}{Listed is the ion and its isoelectronic sequence in parentheses}
\tablenotetext{b}{blends with a Li-like transition}
\tablenotetext{c}{blends with lower charge states}

\tablecomments{Listed are the statistical uncertainties, which are in
  addition to 0.13\,eV (Si) and 0.23\,eV (S) systematic uncertainty. A
  fit to the FAC models (given uncertainties derived from the fit),
  the energy of the strongest line according to \citet{behar02a}, and
  the lines by \citet{house69a} averaging over the fine structure are
  listed as well. For O-like Si and S, \citet{behar02a} and
  \citet{house69a} list exactly the same value to three decimals in units of
  \AA .}
\end{deluxetable*}

\begin{deluxetable*}{lccccccc}
\tablecaption{Doppler shifts in $\mathrm{km\,s}^{-1}$ for Vela X-1\label{tab:vela}}
\tablecolumns{8}
\tablewidth{0pt}
\tablehead{
 \colhead{Ion} &
 \multicolumn{4}{c}{$\phi=0.0$} & &
 \multicolumn{2}{c}{$\phi=0.5$}\\
 \cline{2-5} \cline{7-8} 
 \colhead{} & \colhead{S02\tablenotemark{a}} & \colhead{G04\tablenotemark{b}} & 
 \colhead{S02 new\tablenotemark{c}} & \colhead{G04 new\tablenotemark{c}} & 
 \colhead{} & \colhead{G04\tablenotemark{b}} & \colhead{G04 new\tablenotemark{c}}}
\startdata
\ion{Si}{9} & $-432\pm173$ & $-570^{+271}_{-490}$ & $383\pm173$ &
$244^{+272}_{-491}$ && $-1028^{+275}_{-137}$ & $-215^{+276}_{-137}$\\
\ion{Si}{8} & $\phantom{-}\phn43\pm214$ & $-119^{+389}_{-488}$ & $479\pm214$ &
$317^{+390}_{-489}$ && $-396$ & $40$\\
\ion{Si}{7} & $-170\pm170$ & $-85$ & $\phn48\pm170$ & $133$ &&
$-527^{+321}_{-249}$ & $-309^{+321}_{-249}$\\
\ion{Si}{6} & $\phantom{-}\phn\phn0\pm211$ & \nodata & $-9\pm211$ &
\nodata && \nodata & \nodata \\
\enddata

\tablenotetext{a}{Doppler shifts from \citet{schulz02a}}
\tablenotetext{b}{Doppler shifts from \citet{goldstein04a}}
\tablenotetext{c}{Doppler shifts from S02 and G04, respectively,
adjusted for the new reference energies measured at EBIT
(Table~\ref{tab:center})}

\end{deluxetable*}

\begin{figure}
\vspace{2mm}
\includegraphics[width=\columnwidth]{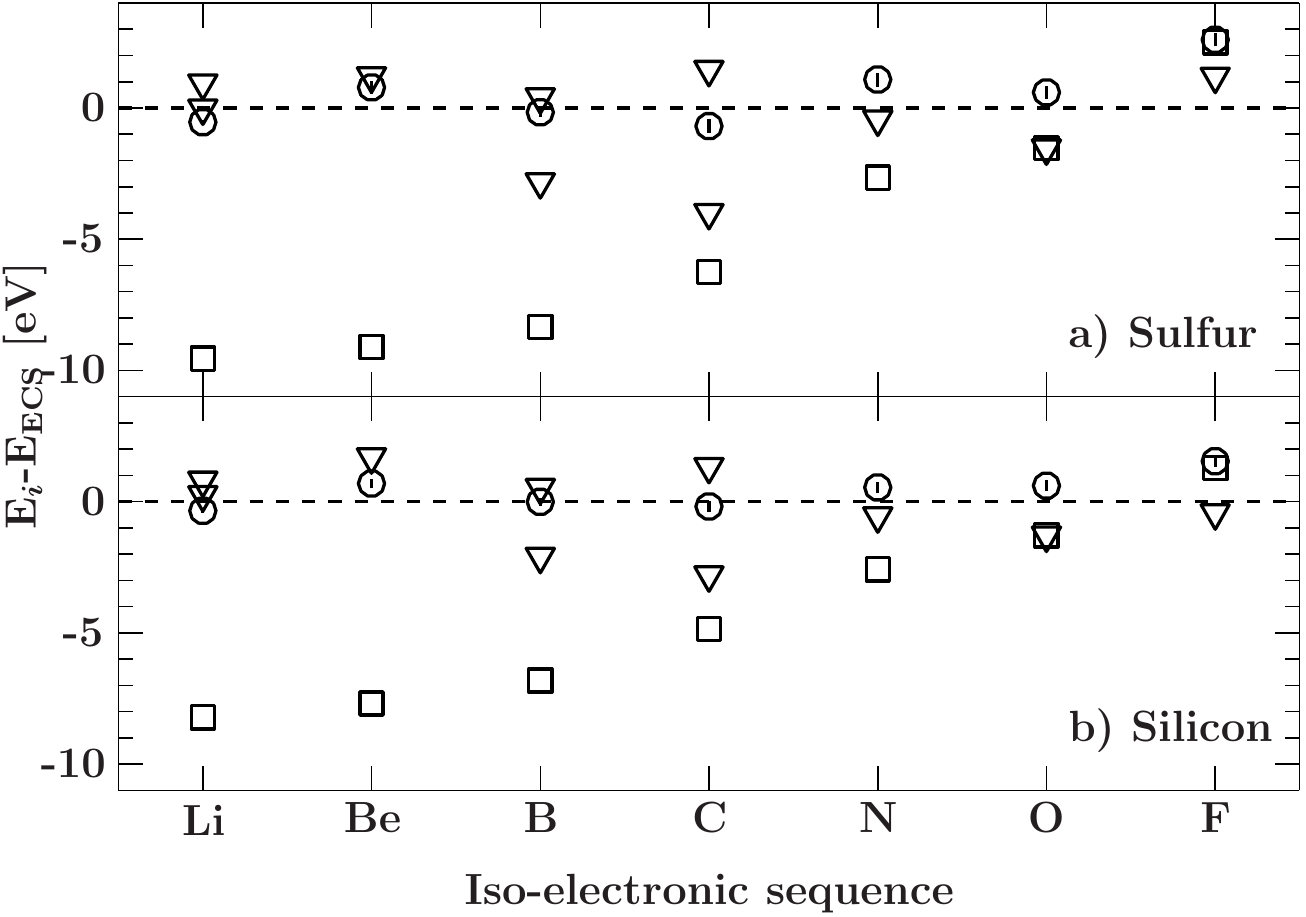}
\caption{Deviation of the theoretical line centers in
  Table~\ref{tab:center} from the centers measured with the ECS, as a
  function of iso-electronic sequence. While the lines derived from
  FAC ({\large$\circ$}) and taken from
  \citet[$\bigtriangledown$]{behar02a} agree fairly well with our
  measurement, the deviation of the \citet[$\square$]{house69a} values
  become significant for higher charge states.}
\label{fig:center}
\end{figure}

\section{Center of Line Blends}\label{sec:blends}

In most experimental cases, the spectral resolution is not adequate to
distinguish between single features of the given lines. This is
especially true for satellite based instrumentation. Therefore, the
ability to determine the energy for each identified line is precluded.
In order to provide the means to derive Doppler shifts also for these
cases, in a second approach we fit the spectra with a single Gaussian
for each of the readily distinguishable line blends, leaving the line
widths free to vary. The obtained line centers, which are listed in
Table~\ref{tab:center} and displayed in Figure~\ref{fig:center}, are
sufficient as reference energies to derive Doppler shifts for
collision dominated or photoionized plasmas where 1s--2p transitions
are dominant, as demonstrated below. Note that again the listed
uncertainties are on the 90\% confidence level and in addition to a
systematic uncertainty of 0.13\,eV for Si and 0.23\,eV for S. For
comparison, we also fitted our FAC models with Gaussians\footnote{Note
  that the uncertainties on these values are derived from the fit and
  purely statistical}, and list the reference energies from
\citet{behar02a} and \citet{house69a} in Table~\ref{tab:center}. Based
on a similar argument, \citet{behar02a} only list the strongest
(photo-absorption) lines, i.e., lines with the largest oscillator
strength, for the isoelectronic sequences \ion{He}{1} to \ion{F}{1} of
the most common elements in astrophysics. According to their
calculations, these lines account for more than 70\%, and in most
cases even more than 90\%, of the absorption effect for the respective
transitions. The good agreement of the energies of their principal
lines with our measurements supports their argument. The
\citet{house69a} tables, on the other hand, deviate significantly from
our results, especially for the higher charge states; this is probably
a direct consequence of averaging over the fine structure.

\begin{figure}[t!]
\includegraphics[width=\columnwidth]{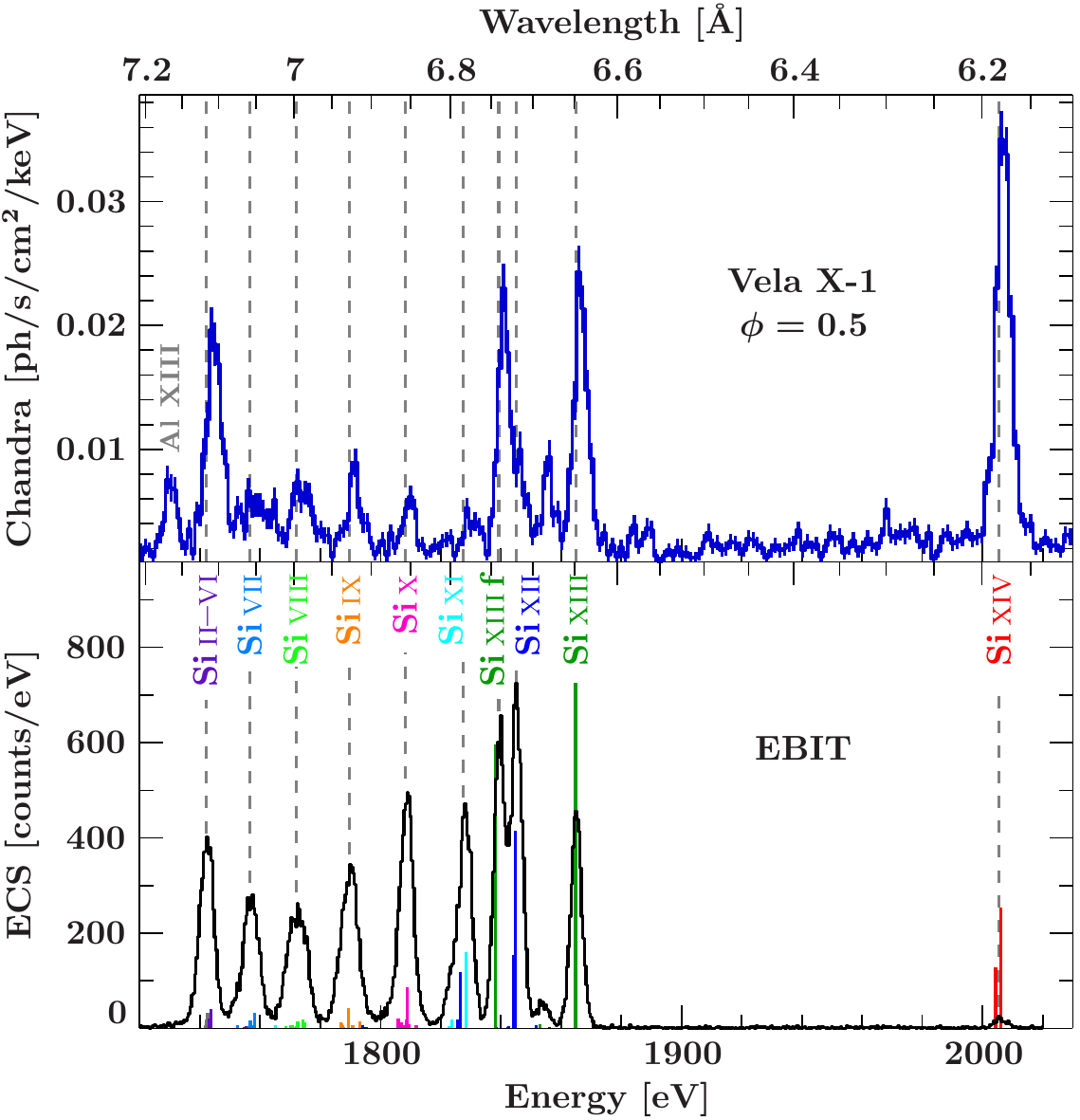}
\caption{Comparison of the Vela X-1 Si spectral region observed by
\textsl{Chandra}-HETG at orbital phase $\phi=0.5$ (ObsID~1927) to the EBIT/ECS spectrum. The
colored sticks are the transitions calculated with FAC.}
\label{fig:vela}
\end{figure}

To demonstrate the impact of our measurements, we use our new
reference energies (Table~\ref{tab:center}) to re-determine Doppler
shifts for Vela X-1 from the published wavelengths.
Figure~\ref{fig:vela} shows a comparison between our laboratory Si
spectra and the ones measured with \textsl{Chandra}-HETG at orbital phase
$\phi_\mathrm{orb}=0.5$. \citet[][$\phi_\mathrm{orb}=0.0$, i.e.,
eclipse]{schulz02a} and \citet[][$\phi_\mathrm{orb}=0$ and
$\phi_\mathrm{orb}=0.5$]{goldstein04a} both fitted the lines
originating from some of the lower charge states of Si in the \textsl{Chandra}
spectra, but did not model the intermediate charge states up to
Li-like. Using \citet{house69a} as a reference for the transition
wavelengths resulted in Doppler shifts that not only differ between
these charge states, but also deviate significantly from the He- and
H-like ions in the same observation, even switching signs from blue to
red-shifted \citep{goldstein04a}. Determining the Doppler shifts based
on our laboratory reference spectra (Table~\ref{tab:vela}) results in
Doppler shifts that are similar for all Si ions and now also agree
with the Doppler shifts determined from the He- and H-like species,
for which the rest-wavelengths are well known. This is more consistent
with the picture of photons being reprocessed by clumps of material
with an onion-like structure, where the outer layers shield the colder
and denser core of the clump from the ionizing radiation of the
compact object.  Similar results were also obtained for Cyg X-1
\citep{hell2013a}, where these lines are seen in absorption.

\section{Summary}\label{sec:conclusion} 

The K$\alpha$ emission line energies from Si$^{4+}$ through Si$^{12+}$ and
S$^{6+}$ through S$^{14+}$ have been measured using the ECS
calorimeter at the LLNL EBIT facility. The results have been compared
to our own FAC calculations and earlier calculations of
\citet{palmeri08a}, \citet{behar02a}, and \citet{house69a}. The newly
available data (Table~\ref{tab:center}) can directly be applied to
resolve astrophysical problems such as, e.g., wind diagnostics in high
mass X-ray binary systems like Vela X-1 \citep{liedahl08a} and Cyg X-1
\citep{miskovicova:16a}. The 90\% confidence limits of
$\lesssim0.5$\,eV on the measured line centers presented here
correspond to Doppler shifts of less than 90\,$\mathrm{km\,s}^{-1}$.
These measurements, therefore, provide line centers with an accuracy
slightly better than the uncertainty of $\sim100\,\mathrm{km\,s}^{-1}$
on the \textsl{Chandra} HETG
\citep{marshall2004a,canizares05a,chandrapog2015}.
When future missions with higher effective area make high-resolution
spectra of point as well as extended celestial sources more commonly 
available, we expect to see these lines to be resolved in a variety of 
sources. Our results will
then be especially useful for extended sources like supernova
remnants which have yet to be observed in high resolution.

\acknowledgements This work was performed by LLNL under the auspices
of the US DOE under Contract DE-AC52-07NA27344. It was supported by
NASA work orders NNX/2AH84G; by the Bundesministerium f\"ur Wirtschaft
und Technologie under grant number \mbox{DLR 50 OR 1113}; by the
European Space Agency under contract No.~4000114313/15/NL/CB; and by
NASA through the Smithsonian Astrophysical Observatory (SAO) contract
SV3-73016 to MIT for support of the \textsl{Chandra} X-ray Center (CXC) and
Science Instruments. CXC is operated by SAO for and on the behalf of
NASA under contract NAS8-03060. This research has made use of ISIS
functions provided by ECAP/Remeis observatory and MIT
(\url{http://www.sternwarte.uni-erlangen.de/isis/}). We thank John E.
Davis for providing the \texttt{slxfig} module used for creating the
presented plots. CHIANTI is a collaborative project involving George
Mason University, the University of Michigan (USA) and the University
of Cambridge (UK).

\software{ISIS \url{http://space.mit.edu/asc/isis/}, ISISscripts
  \url{http://www.sternwarte.uni-erlangen.de/isis/}, FAC v1.1.1}

\appendix
\section{Overview over spectral fits}\label{sec:overview}

\begin{figure*}
\includegraphics[width=0.95\textwidth]{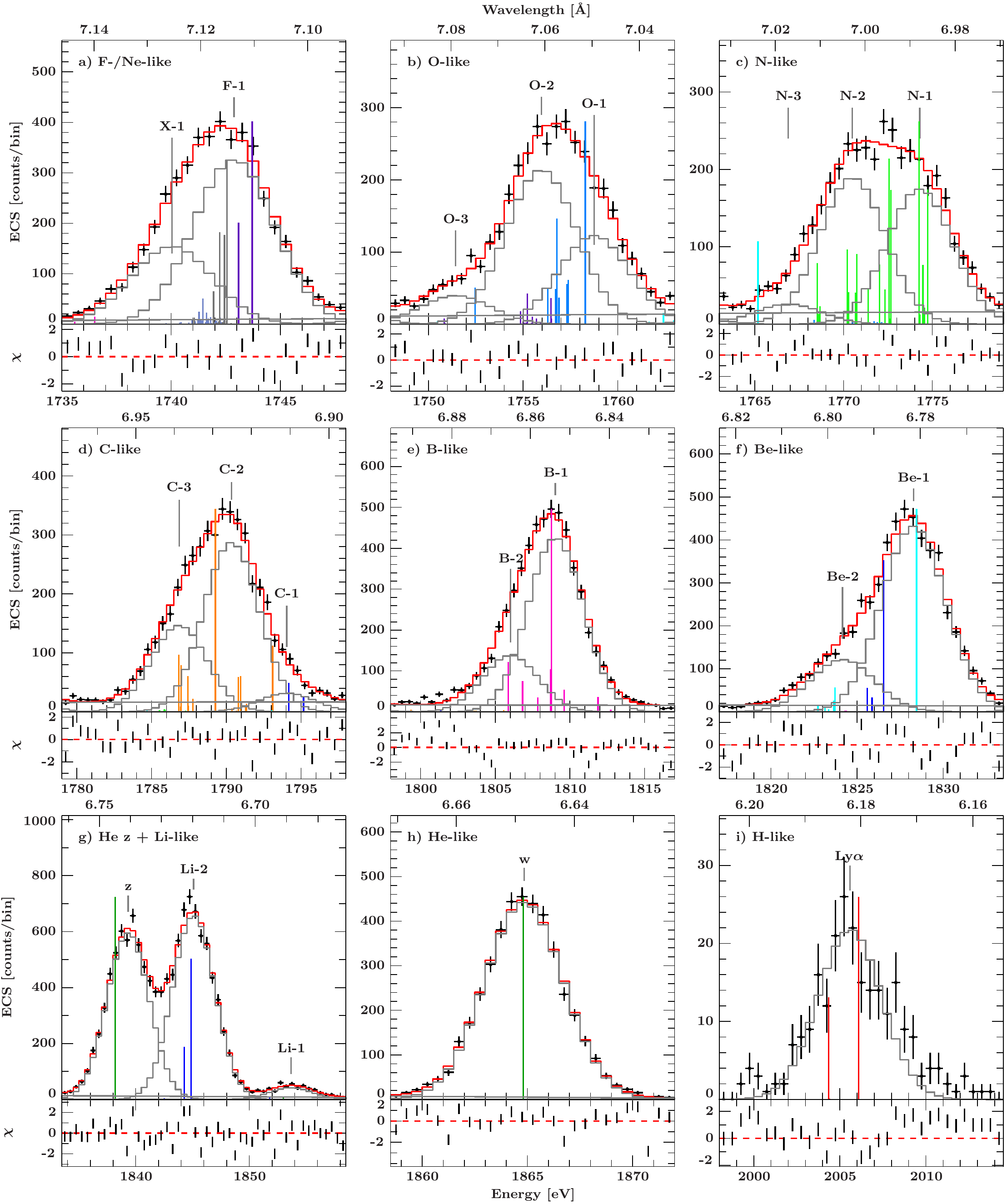} 
\caption{Fit of the measured Si K$\alpha$ spectrum. The data are shown in
  black, the red line shows the model, model components are shown as
  gray lines. Vertical lines represent the theoretical predictions
  according to FAC, color-coded for charge states (see
  Figure~\ref{fig:spectra}). The FAC lines are renormalized such that
  the strongest FAC line in each panel matches the highest peak, i.e.,
  relative FAC line strengths are preserved within but not between
  panels. Line labels can be used as an identifier for the transitions
  listed in Table~\ref{tab:sifit}.} \label{fig:si-fits}
\end{figure*}

\begin{figure*}
\centering
\includegraphics[width=0.95\textwidth]{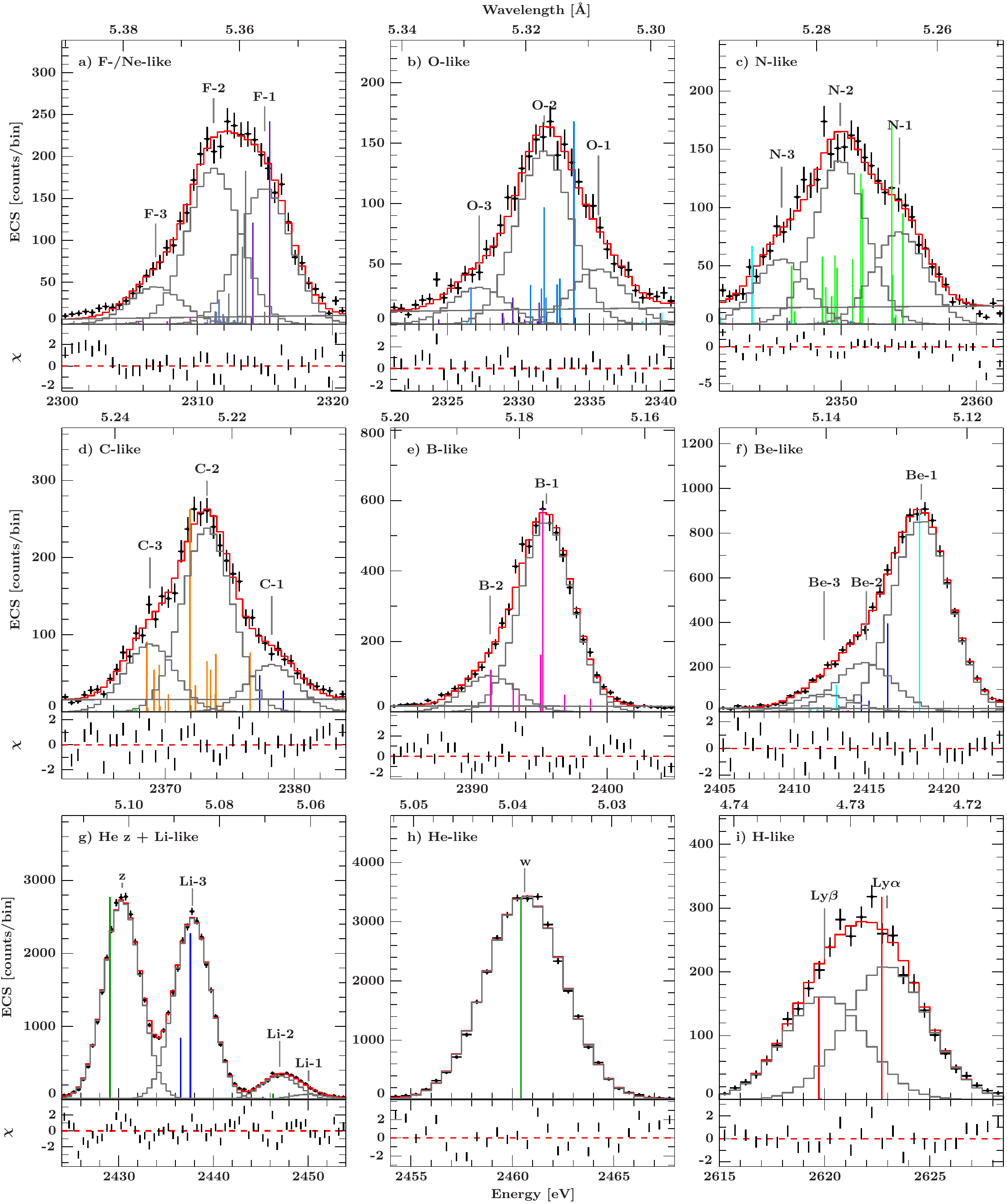}

\caption{Same as Figure~\ref{fig:si-fits} for the S spectrum. Line
  labels denote transitions listed in
  Table~\ref{tab:sfit}.}\label{fig:s-fits}
\end{figure*}

In this appendix we present the full list of measured
line energies obtained with EBIT for Si (Table~\ref{tab:sifit}) and S
(Table~\ref{tab:sfit}). The tables contain the best-fit values from
the EBIT measurement, their identification with transitions from FAC
calculations in $jj$-coupling including the calculated line energy,
and, in $LS$-coupling, a comparison to calculations by P08 and
database entries of CHIANTI, where available.
In addition, Figures~\ref{fig:si-fits} and \ref{fig:s-fits} give a
detailed overview of the data, the best-fit including the individual
Gaussian model components, and the location and theoretical relative
line strengths of the transitions according to the FAC calculations. 

\renewcommand{\thefootnote}{\alph{footnote}}

\setlength{\tabcolsep}{2pt}

\begin{deluxetable*}{lcllllllll}
\tablecolumns{10}
\tabletypesize{\smalltablefont}

\tablecaption{Identification of the fitted silicon line centers.  \label{tab:sifit}}

\tablehead{
  \colhead{} & \colhead{} & \colhead{} & \multicolumn{2}{c}{jj-coupling} & 
  \colhead{} &  \multicolumn{2}{c}{LS-coupling} &\colhead{} & \colhead{} 
  \\\cline{4-5}\cline{7-8}
 \colhead{Key} & \colhead{Fit (eV)} & \colhead{Ion} & \colhead{Lower Level} 
 & \colhead{Upper Level} & \colhead{FAC (eV)} & \colhead{Lower Level} & 
 \colhead{Upper Level} & \colhead{P08 (eV)} & \colhead{CHIANTI}
 }
\startdata
Li-1 & $1853.67\pm0.20$ & He-like Si\,{\sc xiii} & 1\nsf & $(1\ns\,2\np)^{}_1$ & 1852.98 & 1\lss[2]\ \slj{1}{S}{0} & 1\lss2\lsp\ \sljo{3}{P}{1} & 1853.30 & 1853.75\\ 
  && Li-like Si\,{\sc xii} & 1\nsf\,3\ns & $((1\ns\,2\ns)^{}_0\,3\npp)^{}_{3/2}$ & 1851.80 & \nodata & \nodata & \nodata & \nodata\\ 
  && Li-like Si\,{\sc xii} & 1\nsf\,2\npp & $(1\ns\,(2\npp[2])^{}_0)^{}_{1/2}$ & 1856.13 & 1\lss[2]2\lsp\ \sljo{2}{P}{3/2} & 1\lss(\sle{2}{S})\,2\lsp[2](\sle{1}{S})\ \slj{2}{S}{1/2} & 1856.78 & 1854.37\\ 
Li-2 & $1845.09\pm0.05$ & Li-like Si\,{\sc xii} & 1\nsf\,2\ns & $((1\ns\,2\ns)^{}_0\,2\np)^{}_{1/2}$ & 1844.29 & 1\lss[2]2\lss\ \slj{2}{S}{1/2} & 1\lss(\sle{2}{S})\,2\lss2\lsp(\slo{3}{P})\ \sljo{2}{P}{1/2} & 1845.11 & 1843.66\\ 
  && Li-like Si\,{\sc xii} & 1\nsf\,2\ns & $((1\ns\,2\ns)^{}_0\,2\npp)^{}_{3/2}$ & 1844.86 & 1\lss[2]2\lss\ \slj{2}{S}{1/2} & 1\lss(\sle{2}{S})\,2\lss2\lsp(\slo{3}{P})\ \sljo{2}{P}{3/2} & 1845.66 & 1844.21\\ 
z & $1839.33\pm0.05$ & He-like Si\,{\sc xiii} & 1\nsf & $(1\ns\,2\ns)^{}_1$ & 1838.20 & 1\lss[2]\ \slj{1}{S}{0} & 1\lss2\lss\ \slj{3}{S}{1} & \nodata & 1839.42\\ 
Be-1 & $1828.29^{+0.07}_{-0.08}$ & Be-like Si\,{\sc xi} & 1\nsf\,2\nsf & $(1\ns\,2\nsf\,2\npp)^{}_1$ & 1828.46 & 1\lss[2]2\lss[2]\ \slj{1}{S}{0} & 1\lss2\lss[2]2\lsp\ \sljo{1}{P}{1} & 1828.19 & \nodata\\ 
  && Li-like Si\,{\sc xii} & 1\nsf\,2\ns & $((1\ns\,2\ns)^{}_1\,2\npp)^{}_{5/2}$ & 1826.55 & 1\lss[2]2\lss\ \slj{2}{S}{1/2} & 1\lss(\sle{2}{S})\,2\lss2\lsp(\slo{3}{P})\ \sljo{4}{P}{5/2} & \nodata & 1828.19\\ 
Be-2 & $1824.15^{+0.18}_{-0.20}$ & Be-like Si\,{\sc xi} & $1\nsf\,(2\ns\,2\npp)^{}_2$ & $((1\ns\,2\ns)^{}_1\,(2\npp[2])^{}_2)^{}_2$ & 1823.71 & 1\lss[2]2\lss2\lsp\ \sljo{3}{P}{2} & 1\lss(\sle{2}{S})\,2\lss2\lsp[2](\sle{4}{P})\ \slj{3}{P}{2} & 1823.43 & \nodata\\ 
  && Be-like Si\,{\sc xi} & $1\nsf\,(2\ns\,2\np)^{}_0$ & $(((1\ns\,2\ns)^{}_0\,2\np)^{}_{1/2}\,2\npp)^{}_1$ & 1823.64 & 1\lss[2]2\lss2\lsp\ \sljo{3}{P}{0} & 1\lss(\sle{2}{S})\,2\lss[2]2\lsp(\sle{2}{D})\ \slj{3}{D}{1} & 1823.32 & \nodata\\ 
B-1 & $1809.02^{+0.10}_{-0.15}$ & B-like Si\,{\sc x} & 1\nsf\,2\nsf\,2\npp & $(1\ns\,2\nsf\,(2\npp[2])^{}_2)^{}_{3/2}$ & 1808.76 & 1\lss[2]2\lss[2]2\lsp\ \sljo{2}{P}{3/2} & 1\lss2\lss[2]2\lsp[2]\ \slj{2}{P}{3/2} & 1808.38 & \nodata\\ 
  && B-like Si\,{\sc x} & 1\nsf\,2\nsf\,2\np & $((1\ns\,2\nsf\,2\np)^{}_1\,2\npp)^{}_{1/2}$ & 1808.71 & 1\lss[2]2\lss[2]2\lsp\ \sljo{2}{P}{1/2} & 1\lss2\lss[2]2\lsp[2]\ \slj{2}{P}{1/2} & 1808.38 & \nodata\\ 
B-2 & $1806.02^{+0.29}_{-0.49}$ & B-like Si\,{\sc x} & 1\nsf\,2\nsf\,2\npp & $((1\ns\,2\nsf\,2\np)^{}_1\,2\npp)^{}_{5/2}$ & 1805.88 & 1\lss[2]2\lss[2]2\lsp\ \sljo{2}{P}{3/2} & 1\lss2\lss[2]2\lsp[2]\ \slj{2}{D}{5/2} & 1805.32 & \nodata\\ 
  && B-like Si\,{\sc x} & 1\nsf\,2\nsf\,2\np & $((1\ns\,2\nsf\,2\np)^{}_0\,2\npp)^{}_{3/2}$ & 1806.83 & 1\lss[2]2\lss[2]2\lsp\ \sljo{2}{P}{1/2} & 1\lss2\lss[2]2\lsp[2]\ \slj{2}{D}{3/2} & 1806.11 & \nodata\\ 
C-1 & $1794.0\pm1.0$ & C-like Si\,{\sc ix} & $1\nsf\,(2\nsf\,2\np\,2\npp)^{}_2$ & $(1\ns\,2\nsf\,(2\npp[3])^{}_{3/2})^{}_1$ & 1793.10 & 1\lss[2]2\lss[2]2\lsp[2]\ \slj{1}{D}{2} & 1\lss2\lss[2]2\lsp[3]\ \sljo{1}{P}{1} & 1793.23 & \nodata\\ 
  && Li-like Si\,{\sc xii} & 1\nsf\,2\npp & 1\ns\,2\nsf & 1794.18 & 1\lss[2]2\lsp\ \sljo{2}{P}{3/2} & 1\lss2\lss[2]\ \slj{2}{S}{1/2} & 1795.26 & 1794.29\\ 
  && Li-like Si\,{\sc xii} & 1\nsf\,2\np & 1\ns\,2\nsf & 1795.17 & 1\lss[2]2\lsp\ \sljo{2}{P}{1/2} & 1\lss2\lss[2]\ \slj{2}{S}{1/2} & 1796.27 & 1795.31\\ 
C-2 & $1790.34^{+0.25}_{-0.40}$ & C-like Si\,{\sc ix} & $1\nsf\,(2\nsf\,2\np\,2\npp)^{}_2$ & $((1\ns\,2\nsf\,2\np)^{}_1\,(2\npp[2])^{}_2)^{}_2$ & 1789.27 & 1\lss[2]2\lss[2]2\lsp[2]\ \slj{1}{D}{2} & 1\lss2\lss[2]2\lsp[3]\ \sljo{1}{D}{2} & 1789.09 & \nodata\\ 
  && C-like Si\,{\sc ix} & 1\nsf\,2\nsf\,$(2\npp[2])^{}_2$ & $((1\ns\,2\nsf\,2\np)^{}_1\,(2\npp[2])^{}_0)^{}_1$ & 1790.97 & 1\lss[2]2\lss[2]2\lsp[2]\ \slj{3}{P}{2} & 1\lss2\lss[2]2\lsp[3]\ \sljo{3}{P}{1} & 1790.59 & \nodata\\ 
  && C-like Si\,{\sc ix} & 1\nsf\,2\nsf\,$(2\npp[2])^{}_2$ & $(1\ns\,2\nsf\,(2\npp[3])^{}_{3/2})^{}_2$ & 1790.81 & 1\lss[2]2\lss[2]2\lsp[2]\ \slj{3}{P}{2} & 1\lss2\lss[2]2\lsp[3]\ \sljo{2}{P}{2} & 1790.41 & \nodata\\ 
C-3 & $1786.85^{+0.25}_{-0.35}$ & C-like Si\,{\sc ix} & 1\nsf\,2\nsf\,$(2\npp[2])^{}_2$ & $((1\ns\,2\nsf\,2\np)^{}_1\,(2\npp[2])^{}_2)^{}_3$ & 1786.83 & 1\lss[2]2\lss[2]2\lsp[2]\ \slj{3}{P}{2} & 1\lss2\lss[2]2\lsp[3]\ \sljo{3}{D}{3} & 1786.26 & \nodata\\ 
  && C-like Si\,{\sc ix} & 1\nsf\,2\nsf\,$(2\npp[2])^{}_0$ & $(1\ns\,2\nsf\,(2\npp[3])^{}_{3/2})^{}_1$ & 1786.98 & 1\lss[2]2\lss[2]2\lsp[2]\ \slj{1}{S}{0} & 1\lss2\lss[2]2\lsp[3]\ \sljo{1}{P}{1} & 1786.88 & \nodata\\ 
  && C-like Si\,{\sc ix} & $1\nsf\,(2\nsf\,2\np\,2\npp)^{}_1$ & $((1\ns\,2\nsf\,2\np)^{}_0\,(2\npp[2])^{}_2)^{}_2$ & 1787.43 & 1\lss[2]2\lss[2]2\lsp[2]\ \slj{3}{P}{1} & 1\lss2\lss[2]2\lsp[3]\ \sljo{3}{D}{2} & 1786.67 & \nodata\\ 
N-1 & $1774.29^{+0.20}_{-0.19}$ & N-like Si\,{\sc viii} & 1\nsf\,2\nsf\,$(2\np\,(2\npp[2])^{}_2)^{}_{5/2}$ & $((1\ns\,2\nsf\,2\np)^{}_1\,(2\npp[3])^{}_{3/2})^{}_{3/2}$ & 1774.25 & 1\lss[2]2\lss[2]2\lsp[3]\ \sljo{2}{D}{5/2} & 1\lss2\lss[2]2\lsp[4]\ \slj{2}{P}{3/2} & 1773.66 & \nodata\\ 
  && N-like Si\,{\sc viii} & 1\nsf\,2\nsf\,$(2\np\,(2\npp[2])^{}_2)^{}_{3/2}$ & $((1\ns\,2\nsf\,2\np)^{}_1\,(2\npp[3])^{}_{3/2})^{}_{1/2}$ & 1774.74 & 1\lss[2]2\lss[2]2\lsp[3]\ \sljo{2}{D}{3/2} & 1\lss2\lss[2]2\lsp[4]\ \slj{2}{P}{1/2} & 1774.19 & \nodata\\ 
  && N-like Si\,{\sc viii} & 1\nsf\,2\nsf\,$(2\npp[3])^{}_{3/2}$ & $(1\ns\,2\nsf\,2\np[2]\,(2\npp[2])^{}_0)^{}_{1/2}$ & 1774.47 & 1\lss[2]2\lss[2]2\lsp[3]\ \sljo{2}{P}{3/2} & 1\lss2\lss[2]2\lsp[4]\ \slj{2}{S}{1/2} & 1773.86 & \nodata\\ 
N-2 & $1770.5^{+0.5}_{-4.9}$ & N-like Si\,{\sc viii} & 1\nsf\,2\nsf\,$(2\npp[3])^{}_{3/2}$ & $((1\ns\,2\nsf\,2\np)^{}_1\,(2\npp[3])^{}_{3/2})^{}_{3/2}$ & 1770.21 & 1\lss[2]2\lss[2]2\lsp[3]\ \sljo{2}{P}{3/2} & 1\lss2\lss[2]2\lsp[4]\ \slj{2}{P}{3/2} & 1769.46 & \nodata\\ 
  && N-like Si\,{\sc viii} & 1\nsf\,2\nsf\,$(2\np\,(2\npp[2])^{}_0)^{}_{1/2}$ & $((1\ns\,2\nsf\,2\np)^{}_1\,(2\npp[3])^{}_{3/2})^{}_{1/2}$ & 1770.74 & 1\lss[2]2\lss[2]2\lsp[3]\ \sljo{2}{P}{1/2} & 1\lss2\lss[2]2\lsp[4]\ \slj{2}{P}{1/2} & 1770.01 & \nodata\\ 
  && N-like Si\,{\sc viii} & 1\nsf\,2\nsf\,$(2\np\,(2\npp[2])^{}_2)^{}_{5/2}$ & $((1\ns\,2\nsf\,2\np)^{}_1\,(2\npp[3])^{}_{3/2})^{}_{5/2}$ & 1772.56 & 1\lss[2]2\lss[2]2\lsp[3]\ \sljo{2}{D}{5/2} & 1\lss2\lss[2]2\lsp[4]\ \slj{2}{D}{5/2} & 1771.78 & \nodata\\ 
  && N-like Si\,{\sc viii} & 1\nsf\,2\nsf\,$(2\np\,(2\npp[2])^{}_2)^{}_{3/2}$ & $(1\ns\,2\nsf\,2\np[2]\,(2\npp[2])^{}_2)^{}_{3/2}$ & 1172.66 & 1\lss[2]2\lss[2]2\lsp[3]\ \sljo{2}{D}{3/2} & 1\lss2\lss[2]2\lsp[4]\ \slj{2}{D}{3/2} & 1771.76 & \nodata\\ 
  && N-like Si\,{\sc viii} & 1\nsf\,2\nsf\,$(2\np\,(2\npp[2])^{}_2)^{}_{3/2}$ & $((1\ns\,2\nsf\,2\np)^{}_0\,(2\npp[3])^{}_{3/2})^{}_{3/2}$ & 1172.02 & 1\lss[2]2\lss[2]2\lsp[3]\ \sljo{4}{S}{3/2} & 1\lss2\lss[2]2\lsp[4]\ \slj{4}{P}{3/2} & 1770.60 & \nodata\\ 
N-3 & $1766.9^{+1.0}_{-1.3}$ & N-like Si\,{\sc viii} & 1\nsf\,2\nsf\,$(2\npp[3])^{}_{3/2}$ & $((1\ns\,2\nsf\,2\np)^{}_1\,(2\npp[3])^{}_{3/2})^{}_{5/2}$ & 1768.52 & 1\lss[2]2\lss[2]2\lsp[3]\ \sljo{2}{P}{3/2} & 1\lss2\lss[2]2\lsp[4]\ \slj{2}{D}{5/2} & 1767.59 & \nodata\\ 
  && Be-like Si\,{\sc xi} & 1\nsf\,$(2\np\,2\npp)^{}_2$ & $(1\ns\,2\nsf\,2\npp)^{}_1$ & 1765.18 & 1\lss[2]2\lsp[2]\ \slj{1}{D}{2} & 1\lss2\lss[2]2\lsp\ \sljo{1}{P}{1} & 1766.01 & \nodata\\ 
O-1 & $1758.7\pm0.5$ & O-like Si\,{\sc vii} & 1\nsf\,2\nsf\,$(2\np\,(2\npp[3])^{}_{3/2})^{}_2$ & $(1\ns\,2\nsf\,2\np[2]\,(2\npp[3])^{}_{3/2})^{}_1$ & 1758.28 & 1\lss[2]2\lss[2]2\lsp[4]\ \slj{1}{D}{2} & 1\lss2\lss[2]2\lsp[5]\ \sljo{1}{P}{1} & 1756.35 & \nodata\\ 
O-2 & $1756.0\pm0.4$ & O-like Si\,{\sc vii} & 1\nsf\,2\nsf\,2\np[2]\,$(2\npp[2])^{}_2$ & $(1\ns\,2\nsf\,2\np[2]\,(2\npp[3])^{}_{3/2})^{}_2$ & 1756.79 & 1\lss[2]2\lss[2]2\lsp[4]\ \slj{3}{P}{2} & 1\lss2\lss[2]2\lsp[5]\ \sljo{3}{P}{2} & 1754.39 & \nodata\\ 
  && O-like Si\,{\sc vii} & 1\nsf\,2\nsf\,2\np[2]\,$(2\npp[2])^{}_2$ & $(1\ns\,2\nsf\,2\np\,2\npp[4])^{}_1$ & 1757.38 & 1\lss[2]2\lss[2]2\lsp[4]\ \slj{3}{P}{0} & 1\lss2\lss[2]2\lsp[5]\ \sljo{3}{P}{1} & 1754.96 & \nodata\\ 
  && O-like Si\,{\sc vii} & 1\nsf\,2\nsf\,$(2\np\,(2\npp[3])^{}_{3/2})^{}_1$ & $(1\ns\,2\nsf\,2\np\,2\npp[4])^{}_0$ & 1757.32 & 1\lss[2]2\lss[2]2\lsp[4]\ \slj{3}{P}{1} & 1\lss2\lss[2]2\lsp[5]\ \sljo{3}{P}{0} & 1754.78 & \nodata\\ 
  && O-like Si\,{\sc vii} & 1\nsf\,2\nsf\,$(2\np\,(2\npp[3])^{}_{3/2})^{}_1$ & $(1\ns\,2\nsf\,2\np[2]\,(2\npp[3])^{}_{3/2})^{}_2$ & 1756.30 & 1\lss[2]2\lss[2]2\lsp[4]\ \slj{3}{P}{1} & 1\lss2\lss[2]2\lsp[5]\ \sljo{3}{P}{2} & 1753.91 & \nodata\\ 
  && O-like Si\,{\sc vii} & 1\nsf\,2\nsf\,2\npp[4] & $(1\ns\,2\nsf\,2\np\,2\npp[4])^{}_1$ & 1756.70 & 1\lss[2]2\lss[2]2\lsp[4]\ \slj{3}{P}{0} & 1\lss2\lss[2]2\lsp[5]\ \sljo{3}{P}{1} & 1754.29 & \nodata\\ 
O-3 & $1751.4\pm0.6$ &  O-like Si\,{\sc vii} & 1\nsf\,2\nsf\,2\np[2]\,$(2\npp[2])^{}_0$ & $(1\ns\,2\nsf\,2\np[2]\,(2\npp[3])^{}_{3/2})^{}_1$ & 1752.47 & 1\lss[2]2\lss[2]2\lsp[4]\ \slj{1}{S}{0} & 1\lss2\lss[2]2\lsp[5]\ \sljo{1}{P}{1} & 1750.40 & \nodata\\ 
F-1 & $1742.88^{+0.15}_{-0.17}$ & F-like Si\,{\sc vi} & $1\nsf\,2\nsf\,2\np[2]\,(2\npp[3])^{}_{3/2}$ & 1\ns\,2\nsf\,2\npf & 1743.71 & 1\lss[2]2\lss[2]2\lsp[5]\ \sljo{2}{P}{3/2} & 1\lss2\lss[2]2\lsp[6]\ \slj{2}{S}{1/2} & 1740.79 & \nodata\\ 
    && F-like Si\,{\sc vi} & 1\nsf\,2\nsf\,2\np\,2\npp[4] & 1\ns\,2\nsf\,2\npf & 1743.09 & 1\lss[2]2\lss[2]2\lsp[5]\ \sljo{2}{P}{1/2} & 1\lss2\lss[2]2\lsp[6]\ \slj{2}{S}{1/2} & 1740.15 & \nodata\\ 
	&& Ne-like Si\,{\sc v} & 1\nsf\,2\nsf\,2\np[2]\,$((2\npp[3])^{}_{3/2}\,3\ns)^{}_1$ & $(1\ns\,2\nsf\,2\npf\,3\ns)^{}_1$ & 1742.23 & 1\lss[2]2\lss[2]2\lsp[5]3\lss\ \sljo{3}{P}{1} & 1\lss2\lss[2]2\lsp[6]3\lss\ \slj{3}{S}{1} & 1743.06 & \nodata\\ 
	&& Ne-like Si\,{\sc v} & 1\nsf\,2\nsf\,$(2\np\,2\npp[4]\,3\ns)^{}_1$ & $(1\ns\,2\nsf\,2\npf\,3\ns)^{}_0$ & 1742.44 & 1\lss[2]2\lss[2]2\lsp[5]3\lss\ \sljo{1}{P}{1} & 1\lss2\lss[2]2\lsp[6]3\lss\ \slj{1}{S}{0} & 1743.33 & \nodata\\ 
	&& Ne-like Si\,{\sc v} & 1\nsf\,2\nsf\,2\np[2]\,$((2\npp[3])^{}_{3/2}\,3\ns)^{}_2$ & $(1\ns\,2\nsf\,2\npf\,3\ns)^{}_1$ & 1742.56 & 1\lss[2]2\lss[2]2\lsp[5]3\lss\ \sljo{3}{P}{2} & 1\lss2\lss[2]2\lsp[6]3\lss\ \slj{3}{S}{1} & 1743.38 & \nodata\\ 
X-1 & $1740.04^{+0.27}_{-0.36}$ & Na--Si-like Si\,{\sc i--iv} & \nodata & \nodata & \nodata & 
\enddata

\tablecomments{Identification of the fitted Si lines with transitions
  of the FAC simulation. The first column is the key to the line
  labels in Figure~\ref{fig:si-fits}, the third column indicates the
  ionization state. \\
  For the He-like lines the key of
  \citet{gabriel72a} is used. Columns 4--6 show the identification
  with FAC lines, columns 7--9 the corresponding transitions from
  \citet{palmeri08a}. Note \\
  that these calculated
  transition wavelengths listed by P08 have been empirically shifted by
  P08 for ions with $3\leq N\leq 9$, where $N$ is the number of
  electrons. Statistical uncertainties \\
  are shown as 90\,\%
  confidence intervals. There is an additional systematic uncertainty
  of 0.13\,eV on all lines.}

\end{deluxetable*}

\clearpage

\begin{deluxetable*}{lcclllllll}

\tablecolumns{10}
\tabletypesize{\smalltablefont}
\tablecaption{Identification of the fitted sulfur line centers.  \label{tab:sfit}}

\tablehead{
  \colhead{} & \colhead{} & \colhead{} & \multicolumn{2}{c}{jj-coupling} & 
  \colhead{} &  \multicolumn{2}{c}{LS-coupling} &\colhead{} & \colhead{} 
  \\\cline{4-5}\cline{7-8}
 \colhead{Key} & \colhead{Fit (eV)} & \colhead{Ion} & \colhead{Lower Level} 
 & \colhead{Upper Level} & \colhead{FAC (eV)} & \colhead{Lower Level} & 
 \colhead{Upper Level} & \colhead{P08 (eV)} & \colhead{CHIANTI}
 }
\startdata
Li-1 & $2450\pm1.0$ & Li-like S\,{\sc xiv} & 1\nsf\,2\npp & $(1\ns\,(2\npp[2])^{}_0)^{}_{1/2}$ & 2449.95 & 1\lss[2]2\lsp\ \sljo{2}{P}{3/2} & 1\lss(\sle{2}{S})\,2\lsp[2](\sle{1}{S})\ \slj{2}{S}{1/2} & 2450.67 & 2449.26\\ 
  && Li-like S\,{\sc xiv} & 1\nsf\,2\np & $(1\ns\,(2\npp[2])^{}_0)^{}_{1/2}$ & 2451.78 & 1\lss[2]2\lsp\ \sljo{2}{P}{1/2} & 1\lss(\sle{2}{S})\,2\lsp[2](\sle{1}{S})\ \slj{2}{S}{1/2} & 2452.51 & 2451.15\\ 
Li-2 & $2447.02^{+0.19}_{-0.27}$ & Li-like S\,{\sc xiv} & 1\nsf\,2\ns & $((1\ns\,2\ns)^{}_1\,2\npp)^{}_{1/2}$ & 2447.65 & 1\lss[2]2\lss\ \slj{2}{S}{1/2} & 1\lss(\sle{2}{S})\,2\lss2\lsp(\slo{1}{P})\ \sljo{2}{P}{1/2} & 2448.01 & 2447.04\\ 
  && He-like S\,{\sc xv} & 1\nsf & $(1\ns\,2\np)^{}_1$ & 2446.32 & 1\lss[2]\ \slj{1}{S}{0} & 1\lss2\lsp\ \sljo{3}{P}{1} &  2446.65 & 2447.14\\ 
Li-3 & $2437.797^{+0.023}_{-0.024}$ & Li-like S\,{\sc xiv} & 1\nsf\,2\ns & $((1\ns\,2\ns)^{}_0\,2\np)^{}_{1/2}$ & 2436.55 & 1\lss[2]2\lss\ \slj{2}{S}{1/2} & 1\lss(\sle{2}{S})\,2\lss2\lsp(\slo{3}{P})\ \sljo{2}{P}{1/2} & 2437.52 & 2437.04\\ 
  && Li-like S\,{\sc xiv} & 1\nsf\,2\ns & $((1\ns\,2\ns)^{}_0\,2\npp)^{}_{3/2}$ & 2437.58 & 1\lss[2]2\lss\ \slj{2}{S}{1/2} & 1\lss(\sle{2}{S})\,2\lss2\lsp(\slo{3}{P})\ \sljo{2}{P}{3/2} & 2438.47 & 2437.99\\ 
z & $2430.380^{+0.024}_{-0.019}$ & He-like S\,{\sc xv} & 1\nsf & $(1\ns\,2\ns)^{}_1$ & 2429.08 & 1\lss[2]\ \slj{1}{S}{0} & 1\lss2\lss\ \slj{3}{S}{1} & \nodata & 2430.35\\ 
Be-1 & $2418.51^{+0.10}_{-0.09}$ & Be-like S\,{\sc xiii} & 1\nsf\,2\nsf & $(1\ns\,2\nsf\,2\npp)^{}_1$ & 2418.38 & 1\lss[2]2\lss[2]\ \slj{1}{S}{0} & 1\lss2\lss[2]2\lsp\ \sljo{1}{P}{1} & 2418.45 & \nodata\\ 
Be-2 & $2414.7^{+1.0}_{-4.0}$ & Li-like S\,{\sc xiv} & 1\nsf\,2\ns & $((1\ns\,2\ns)^{}_1\,2\npp)^{}_{5/2}$ & 2416.26 & 1\lss[2]2\lss\ \slj{2}{S}{1/2} & 1\lss(\sle{2}{S})\,2\lss2\lsp(\slo{3}{P})\ \sljo{4}{P}{5/2} & \nodata & 2416.99\\ 
  && Li-like S\,{\sc xiv} & 1\nsf\,2\ns & $((1\ns\,2\ns)^{}_1\,2\np)^{}_{1/2}$ & 2414.51 & 1\lss[2]2\lss\ \slj{2}{S}{1/2} & 1\lss(\sle{2}{S})\,2\lss2\lsp(\slo{3}{P})\ \sljo{4}{P}{1/2} & 2414.92 & 2415.24\\ 
  && Li-like S\,{\sc xiv} & 1\nsf\,2\ns & $((1\ns\,2\ns)^{}_1\,2\np)^{}_{3/2}$ & 2415.02 & 1\lss[2]2\lss\ \slj{2}{S}{1/2} & 1\lss(\sle{2}{S})\,2\lss2\lsp(\slo{3}{P})\ \sljo{4}{P}{3/2} & 2415.67 & 2415.76\\ 
Be-3 & $2412.0^{+0.8}_{-1.4}$ & Be-like S\,{\sc xiii} & $1\nsf\,(2\ns\,2\npp)^{}_2$ & $((1\ns\,2\ns)^{}_1\,(2\npp[2])^{}_2)^{}_2$ & 2412.83 & 1\lss[2]2\lss2\lsp\ \sljo{3}{P}{2} & 1\lss(\sle{2}{S})\,2\lss2\lsp[2](\sle{4}{P})\ \slj{3}{P}{2} & 2412.89 & \nodata\\ 
  && Be-like S\,{\sc xiii} & 1\nsf\,$(2\ns\,2\np)^{}_0$ & $(((1\ns\,2\ns)^{}_1\,2\np)^{}_{1/2}\,2\npp)^{}_1$ & 2412.82 & 1\lss[2]2\lss2\lsp\ \sljo{3}{P}{0} & 1\lss(\sle{2}{S})\,2\lss2\lsp[2](\sle{2}{D})\ \slj{3}{D}{1} & 2412.75 & \nodata\\ 
B-1 & $2395.51^{+0.06}_{-0.10}$ & B-like S\,{\sc xii} & 1\nsf\,2\nsf\,2\npp & $(1\ns\,2\nsf\,(2\npp[2])^{}_2)^{}_{3/2}$ & 2395.25 & 1\lss[2]2\lss[2]2\lsp\ \sljo{2}{P}{3/2} & 1\lss2\lss[2]2\lsp[2]\ \slj{2}{P}{3/2} & 2394.90 & \nodata\\ 
  && B-like S\,{\sc xii} & 1\nsf\,2\nsf\,2\np & $((1\ns\,2\nsf\,2\np)^{}_1\,2\npp)^{}_{1/2}$ & 2395.11 & 1\lss[2]2\lss[2]2\lsp\ \sljo{2}{P}{1/2} & 1\lss2\lss[2]2\lsp[2]\ \slj{2}{P}{1/2} & 2394.86 & \nodata\\ 
  && B-like S\,{\sc xii} & 1\nsf\,2\nsf\,2\np & $(1\ns\,2\nsf\,(2\npp[2])^{}_{2})^{}_{1/2}$ & 2396.87 & 1\lss[2]2\lss[2]2\lsp\ \sljo{2}{P}{1/2} & 1\lss2\lss[2]2\lsp[2]\ \slj{2}{P}{3/2} & 2396.52 & \nodata\\ 
B-2 & $2391.36^{+0.20}_{-0.42}$ & B-like S\,{\sc xii} & 1\nsf\,2\nsf\,2\npp & $((1\ns\,2\nsf\,2\np)^{}_1\,2\npp)^{}_{5/2}$ & 2391.41 & 1\lss[2]2\lss[2]2\lsp\ \sljo{2}{P}{3/2} & 1\lss2\lss[2]2\lsp[2]\ \slj{2}{D}{5/2} & 2390.87 & \nodata\\ 
  && B-like S\,{\sc xii} & 1\nsf\,2\nsf\,2\np &  $((1\ns\,2\nsf\,2\np)^{}_0\,2\npp)^{}_{3/2}$ & 2393.07 & 1\lss[2]2\lss[2]2\lsp\ \sljo{2}{P}{1/2} & 1\lss2\lss[2]2\lsp[2]\ \slj{2}{D}{3/2} & 2392.27 & \nodata\\ 
  && B-like S\,{\sc xii} & 1\nsf\,2\nsf\,2\npp & $((1\ns\,2\nsf\,2\np)^{}_1\,2\npp)^{}_{1/2}$ & 2393.50 & 1\lss[2]2\lss[2]2\lsp\ \sljo{2}{P}{3/2} & 1\lss2\lss[2]2\lsp[2]\ \slj{2}{P}{1/2} & 2393.24 & \nodata\\ 
C-1 & $2378.26^{+0.27}_{-0.20}$ & C-like S\,{\sc xi} & 1\nsf\,2\nsf\,$(2\np\,2\npp)^{}_2$ & $(1\ns\,2\nsf\,(2\npp[3])^{}_{3/2})^{}_1$ & 2376.60 & 1\lss[2]2\lss[2]2\lsp[2]\ \slj{1}{D}{2} & 1\lss2\lss[2]2\lsp[3]\ \sljo{1}{P}{1} & 2376.59 & \nodata\\ 
  && Li-like S\,{\sc xiv} & 1\nsf\,2\npp & 1\ns\,2\nsf & 2377.32 & 1\lss[2]2\lsp\ \sljo{2}{P}{3/2} & 1\lss2\lss[2]\ \slj{2}{S}{1/2} & 2378.50 & 2378.32\\ 
  && Li-like S\,{\sc xiv} & 1\nsf\,2\np & 1\ns\,2\nsf & 2379.16 & 1\lss[2]2\lsp\ \sljo{2}{P}{1/2} & 1\lss2\lss[2]\ \slj{2}{S}{1/2} & 2380.37 & 2380.19\\ 
C-2 & $2373.25^{+0.14}_{-0.16}$ & C-like S\,{\sc xi} & 1\nsf\,2\nsf\,$(2\np\,2\npp)^{}_2$ & $((1\ns\,2\nsf\,2\np)^{}_1\,(2\npp[2])^{}_2)^{}_2$ & 2371.92 & 1\lss[2]2\lss[2]2\lsp[2]\ \slj{1}{D}{2} & 1\lss2\lss[2]2\lsp[3]\ \sljo{1}{D}{2} & 2371.59 & \nodata\\ 
  && C-like S\,{\sc xi} & 1\nsf\,2\nsf\,$(2\npp[2])^{}_2$ & $((1\ns\,2\nsf\,2\np)^{}_1\,(2\npp[2])^{}_0)^{}_1$ & 2373.92 & 1\lss[2]2\lss[2]2\lsp[2]\ \slj{3}{P}{2} & 1\lss2\lss[2]2\lsp[3]\ \sljo{3}{P}{1} & 2373.36 & \nodata\\ 
  && C-like S\,{\sc xi} & 1\nsf\,2\nsf\,$(2\np\,2\npp)^{}_1$ & $((1\ns\,2\nsf\,2\np)^{}_1\,(2\npp[2])^{}_2)^{}_1$ & 2373.23 & 1\lss[2]2\lss[2]2\lsp[2]\ \slj{3}{P}{1} & 1\lss2\lss[2]2\lsp[3]\ \sljo{3}{S}{1} & 2372.77 & \nodata\\ 
  && C-like S\,{\sc xi} & 1\nsf\,2\nsf\,$(2\npp[2])^{}_2$ & $(1\ns\,2\nsf\,(2\npp[3])^{}_{3/2})^{}_2$ & 2373.52 & 1\lss[2]2\lss[2]2\lsp[2]\ \slj{3}{P}{2} & 1\lss2\lss[2]2\lsp[3]\ \sljo{3}{P}{2} & 2372.95 & \nodata\\ 
C-3 & $2368.83^{+0.20}_{-0.24}$ & C-like S\,{\sc xi} & 1\nsf\,2\nsf\,$(2\npp[2])^{}_2$ & $((1\ns\,2\nsf\,2\np)^{}_1\,(2\npp[2])^{}_2)^{}_3$ & 2368.58 & 1\lss[2]2\lss[2]2\lsp[2]\ \slj{3}{P}{2} & 1\lss2\lss[2]2\lsp[3]\ \sljo{3}{D}{3} & 2367.83 & \nodata\\ 
  && C-like S\,{\sc xi} & 1\nsf\,2\nsf\,$(2\np\,2\npp)^{}_1$ & $((1\ns\,2\nsf\,2\np)^{}_0\,(2\npp[2])^{}_2)^{}_2$ & 2369.55 & 1\lss[2]2\lss[2]2\lsp[2]\ \slj{3}{P}{1} & 1\lss2\lss[2]2\lsp[3]\ \sljo{3}{D}{2} & 2368.55 & \nodata\\ 
  && C-like S\,{\sc xi} & 1\nsf\,2\nsf\,$(2\npp[2])^{}_0$ & $(1\ns\,2\nsf\,(2\npp[3])^{}_{3/2})^{}_1$ & 2369.14 & 1\lss[2]2\lss[2]2\lsp[2]\ \slj{1}{S}{0} & 1\lss2\lss[2]2\lsp[3]\ \sljo{1}{P}{1} & 2368.87 & \nodata\\ 
N-1 & $2354.33^{+0.23}_{-0.29}$ & N-like S\,{\sc x} & 1\nsf\,2\nsf\,$(2\np\,(2\npp[2])^{}_2)^{}_{5/2}$ & $((1\ns\,2\nsf\,2\np)^{}_1\,(2\npp[3])^{}_{3/2})^{}_{3/2}$ & 2353.74 & 1\lss[2]2\lss[2]2\lsp[3]\ \sljo{2}{D}{5/2} & 1\lss2\lss[2]2\lsp[4]\ \slj{2}{P}{3/2} & 2352.86 & \nodata\\ 
  && N-like S\,{\sc x} & 1\nsf\,2\nsf\,$(2\np\,(2\npp[2])^{}_2)^{}_{3/2}$ & $((1\ns\,2\nsf\,2\np)^{}_1\,(2\npp[3])^{}_{3/2})^{}_{1/2}$ & 2354.57 & 1\lss[2]2\lss[2]2\lsp[3]\ \sljo{2}{D}{3/2} & 1\lss2\lss[2]2\lsp[4]\ \slj{2}{P}{1/2} & 2353.85 & \nodata\\ 
  && N-like S\,{\sc x} & 1\nsf\,2\nsf\,$(2\npp[3])^{}_{3/2}$ & 1\ns\,2\nsf\,2\npp[4] & 2353.80 & 1\lss[2]2\lss[2]2\lsp[3]\ \sljo{2}{P}{3/2} & 1\lss2\lss[2]2\lsp[4]\ \slj{2}{S}{1/2} & 2352.91 & \nodata\\ 
N-2 & $2349.94^{+0.23}_{-0.32}$ & N-like S\,{\sc x} & 1\nsf\,2\nsf\,$(2\np\,(2\npp[2])^{}_2)^{}_{5/2}$ & $((1\ns\,2\nsf\,2\np)^{}_1\,(2\npp[3])^{}_{3/2})^{}_{5/2}$ & 2351.45 & 1\lss[2]2\lss[2]2\lsp[3]\ \sljo{2}{D}{5/2} & 1\lss2\lss[2]2\lsp[4]\ \slj{2}{D}{5/2} & 2350.45 & \nodata\\ 
  && N-like S\,{\sc x} & 1\nsf\,2\nsf\,$(2\np\,(2\npp[2])^{}_2)^{}_{3/2}$ & $(1\ns\,2\nsf\,2\np[2]\,(2\npp[2])^{}_2)^{}_{3/2}$ & 2351.61 & 1\lss[2]2\lss[2]2\lsp[3]\ \sljo{2}{D}{3/2} & 1\lss2\lss[2]2\lsp[4]\ \slj{2}{D}{3/2} & 2350.37 & \nodata\\ 
  && N-like S\,{\sc x} & 1\nsf\,2\nsf\,$(2\np\,(2\npp[2])^{}_0)^{}_{1/2}$ & $((1\ns\,2\nsf\,2\np)^{}_1\,(2\npp[3])^{}_{3/2})^{}_{1/2}$ & 2349.55 & 1\lss[2]2\lss[2]2\lsp[3]\ \sljo{2}{P}{1/2} & 1\lss2\lss[2]2\lsp[4]\ \slj{2}{P}{1/2} & 2348.59 & \nodata\\ 
  && N-like S\,{\sc x} & 1\nsf\,2\nsf\,$(2\npp[3])^{}_{3/2}$ & $((1\ns\,2\nsf\,2\np)^{}_1\,(2\npp[3])^{}_{3/2})^{}_{3/2}$ & 2348.63 & 1\lss[2]2\lss[2]2\lsp[3]\ \sljo{2}{P}{3/2} & 1\lss2\lss[2]2\lsp[4]\ \slj{2}{P}{3/2} & 2347.61 & \nodata\\ 
  && N-like S\,{\sc x} & 1\nsf\,2\nsf\,$(2\np\,(2\npp[2])^{}_2)^{}_{3/2}$ & $((1\ns\,2\nsf\,2\np)^{}_0\,(2\npp[3])^{}_{3/2})^{}_{3/2}$ & 2350.86 & 1\lss[2]2\lss[2]2\lsp[3]\ \sljo{4}{S}{3/2} & 1\lss2\lss[2]2\lsp[4]\ \slj{4}{P}{3/2} & 2349.12 & \nodata\\ 
  && N-like S\,{\sc x} & 1\nsf\,2\nsf\,$(2\np\,(2\npp[2])^{}_2)^{}_{3/2}$ & $(1\ns\,2\nsf\,2\np[2]\,(2\npp[2])^{}_2)^{}_{5/2}$ & 2349.70 & 1\lss[2]2\lss[2]2\lsp[3]\ \sljo{4}{S}{3/2} & 1\lss2\lss[2]2\lsp[4]\ \slj{4}{P}{5/2} & 2348.10 & \nodata\\ 
N-3 & $2345.6 ^{+0.4}_{-0.6}$ & N-like S\,{\sc x} & 1\nsf\,2\nsf\,$(2\npp[3])^{}_{3/2}$ & $((1\ns\,2\nsf\,2\np)^{}_1\,(2\npp[3])^{}_{3/2})^{}_{5/2}$ & 2346.34 & 1\lss[2]2\lss[2]2\lsp[3]\ \sljo{2}{P}{3/2} & 1\lss2\lss[2]2\lsp[4]\ \slj{2}{D}{5/2} & 2345.17 & \nodata\\ 
  && Be-like S\,{\sc xiii} & 1\nsf\,$(2\np\,2\npp)^{}_2$ & $(1\ns\,2\nsf\,2\npp)^{}_1$ & 2343.42 & 1\lss[2]2\lsp[2]\ \slj{1}{D}{2} & 1\lss2\lss[2]2\lsp\ \sljo{1}{P}{1} & 2344.54 & \nodata\\ 
O-1 & $2335.6^{+0.5}_{-4.3}$ & O-like S\,{\sc ix} & 1\nsf\,2\nsf\,$(2\np\,(2\npp[3])^{}_{3/2})^{}_2$ & $(1\ns\,2\nsf\,2\np[2]\,(2\npp[3])^{}_{3/2})^{}_1$ & 2333.91 & 1\lss[2]2\lss[2]2\lsp[4]\ \slj{1}{D}{2} & 1\lss2\lss[2]2\lsp[5]\ \sljo{1}{P}{1} & 2331.76 & \nodata\\ 
O-2 & $2331.82^{+0.27}_{-0.48}$ & O-like S\,{\sc ix} & 1\nsf\,2\nsf\,2\np[2]\,$(2\npp[2])^{}_2$ & $(1\ns\,2\nsf\,2\np[2]\,(2\npp[3])^{}_{3/2})^{}_2$ & 2331.82 & 1\lss[2]2\lss[2]2\lsp[4]\ \slj{3}{P}{2} & 1\lss2\lss[2]2\lsp[5]\ \sljo{3}{P}{2} & 2329.13 & \nodata\\ 
  && O-like S\,{\sc ix} & 1\nsf\,2\nsf\,2\np[2]\,$(2\npp[2])^{}_2$ & $(1\ns\,2\nsf\,2\np\,2\npp[4])^{}_1$ & 2332.91 & 1\lss[2]2\lss[2]2\lsp[4]\ \slj{3}{P}{2} & 1\lss2\lss[2]2\lsp[5]\ \sljo{3}{P}{1} & 2330.18 & \nodata\\ 
  && O-like S\,{\sc ix} & 1\nsf\,2\nsf\,$(2\np\,(2\npp[3])^{}_{3/2})^{}_1$ & $(1\ns\,2\nsf\,2\np\,2\npp[4])^{}_0$ & 2332.76 & 1\lss[2]2\lss[2]2\lsp[4]\ \slj{3}{P}{1} & 1\lss2\lss[2]2\lsp[5]\ \sljo{3}{P}{0} & 2329.83 & \nodata\\ 
  && O-like S\,{\sc ix} & 1\nsf\,2\nsf\,$(2\np\,(2\npp[3])^{}_{3/2})^{}_1$ & $(1\ns\,2\nsf\,2\np[2]\,(2\npp[3])^{}_{3/2})^{}_2$ & 2330.84 & 1\lss[2]2\lss[2]2\lsp[4]\ \slj{3}{P}{1} & 1\lss2\lss[2]2\lsp[5]\ \sljo{3}{P}{2} & 2328.21 & \nodata\\ 
  && O-like S\,{\sc ix} & 1\nsf\,2\nsf\,2\npp[4] & $(1\ns\,2\nsf\,2\np\,2\npp[4])^{}_1$ & 2331.61 & 1\lss[2]2\lss[2]2\lsp[4]\ \slj{3}{P}{0} & 1\lss2\lss[2]2\lsp[5]\ \sljo{3}{P}{1} & 2328.91 & \nodata\\ 
O-3 & $2327.2^{+0.5}_{-0.7}$ & O-like S\,{\sc ix} & 1\nsf\,2\nsf\,2\np[2]\,$(2\npp[2])^{}_0$ & $(1\ns\,2\nsf\,2\np[2]\,(2\npp[3])^{}_{3/2})^{}_1$ & 2316.67 & 1\lss[2]2\lss[2]2\lsp[4]\ \slj{1}{S}{0} & 1\lss2\lss[2]2\lsp[5]\ \sljo{1}{P}{1} & 2324.33 & \nodata\\ 
F-1 & $2315.00^{+0.17}_{-0.24}$ & F-like S\,{\sc viii} & 1\nsf\,2\nsf\,2\np[2]\,$(2\npp[3])^{}_{3/2}$ & 1\ns\,2\nsf\,2\npf & 2315.36 & 1\lss[2]2\lss[2]2\lsp[5]\ \sljo{2}{P}{3/2} & 1\lss2\lss[2]2\lsp[6]\ \slj{2}{S}{1/2} & 2312.40 & \nodata\\ 
  && F-like S\,{\sc viii} & 1\nsf\,2\nsf\,2\np\,2\npp[4] & 1\ns\,2\nsf\,2\npf & 2314.13 & 1\lss[2]2\lss[2]2\lsp[5]\ \sljo{2}{P}{1/2} & 1\lss2\lss[2]2\lsp[6]\ \slj{2}{S}{1/2} & 2311.15 & \nodata\\ 
  && Ne-like S\,{\sc vii} & 1\nsf\,2\nsf\,2\np[2]\,$((2\npp[3])^{}_{3/2}\,3\ns)^{}_2$ & $(1\ns\,2\nsf\,2\np[2]\,2\npp[4]\,3\ns)^{}_1$ & 2313.58 & 1\lss[2]2\lss[2]2\lsp[5]3\lss\ \sljo{3}{P}{2} & 1\lss2\lss[2]2\lsp[6]3\lss\ \slj{3}{S}{1} & 2314.74 & \nodata\\ 
  && Ne-like S\,{\sc vii} & 1\nsf\,2\nsf\,2\np[2]\,$((2\npp[3])^{}_{3/2}\,3\ns)^{}_1$ & $(1\ns\,2\nsf\,2\np[2]\,2\npp[4]\,3\ns)^{}_1$ & 2312.99 & 1\lss[2]2\lss[2]2\lsp[5]3\lss\ \sljo{3}{P}{1} & 1\lss2\lss[2]2\lsp[6]3\lss\ \slj{3}{S}{1} & 2314.17 & \nodata\\ 
  && Ne-like S\,{\sc vii} & 1\nsf\,2\nsf\,$(2\np\,2\npp[4]\,3\ns)^{}_1$ & $(1\ns\,2\nsf\,2\np[2]\,2\npp[4]\,3\ns)^{}_0$ & 2313.35 & 1\lss[2]2\lss[2]2\lsp[5]3\lss\ \sljo{1}{P}{1} & 1\lss2\lss[2]2\lsp[6]3\lss\ \slj{1}{S}{0} & 2312.62 & \nodata\\ 
F-2 & $2311.22^{+0.27}_{+0.41}$ & Na-like S\,{\sc vi} & 1\nsf\,2\nsf\,$((2\np\,2\npp[4]\,3\ns)^{}_1\,3\npp)^{}_{5/2}$ & $((1\ns\,2\nsf\,2\np[2]\,2\npp[4]\,3\ns)^{}_0\,3\npp)^{}_{3/2}$ & 2311.62 & 1\lss[2]2\lss[2]2\lsp[5]3\lss3\lsp\ \slj{2}{D}{5/2} & 1\lss2\lss[2]2\lsp[6]3\lss3\lsp\ \sljo{2}{P}{3/2} & 2312.71 & \nodata\\ 
  && B-like S\,{\sc xii} & 1\nsf\,$(2\np\,(2\npp[2])^{}_2)^{}_{5/2}$ & $(1\ns\,2\nsf\,(2\npp[2])^{}_2)^{}_{3/2}$ & 2309.50 & 1\lss[2]2\lsp[3]\ \sljo{2}{D}{5/2} & 1\lss2\lss[2]2\lsp[2]\ \slj{2}{P}{3/2} & 2310.12 & \nodata\\ 
  && Na--S-like S\,{\sc i--vi} & \nodata & \nodata & \nodata & \\
F-3 & $2306.9^{+0.4}_{-0.7}$ & B-like S\,{\sc xii} & 1\nsf\,$(2\np\,(2\npp[2])^{}_2)^{}_{3/2}$ & $((1\ns\,2\nsf\,2\np)^{}_1\,2\npp)^{}_{1/2}$ & 2307.79 & 1\lss[2]2\lsp[3]\ \sljo{2}{D}{3/2} & 1\lss2\lss[2]2\lsp[2]\ \slj{2}{P}{1/2} & 2308.57 & \nodata\\ 
  && B-like S\,{\sc xii} & 1\nsf\,$(2\np\,(2\npp[2])^{}_2)^{}_{5/2}$ & $((1\ns\,2\nsf\,2\np)^{}_1\,2\npp)^{}_{5/2}$ & 2305.65 & 1\lss[2]2\lsp[3]\ \sljo{2}{D}{5/2} & 1\lss2\lss[2]2\lsp[2]\ \slj{2}{D}{5/2} & 2306.12 & \nodata\\ 
  && B-like S\,{\sc xii} & 1\nsf\,$(2\np\,(2\npp[2])^{}_2)^{}_{3/2}$ & $(1\ns\,2\nsf\,(2\npp[2])^{}_2)^{}_{5/2}$ & 2304.74 & 1\lss[2]2\lsp[3]\ \sljo{4}{S}{3/2} & 1\lss2\lss[2]2\lsp[2]\ \slj{4}{P}{5/2} & 2303.85 & \nodata 
\enddata

\tablecomments{Identification of the fitted S lines with transitions
  of the FAC simulation. The first column is the key to the line
  labels in Figure~\ref{fig:si-fits}, the third column indicates the
  ionization state. For the He-like lines the key of
  \citet{gabriel72a} is used. Columns 4--6 show the identification
  with FAC lines, columns 7--9 the corresponding transitions from
  \citet{palmeri08a}. Note that these calculated
  transition wavelengths listed by P08 have been empirically shifted by
  P08 for ions with $3\leq N\leq 9$, where $N$ is the number of
  electrons. Statistical uncertainties are given as 90\% confidence
  intervals. There is an additional systematic uncertainty of 0.23\,eV
  on all lines.}

\end{deluxetable*}

\clearpage

\renewcommand{\thefootnote}{\arabic{footnote}}

\clearpage

\end{document}